\documentclass[11pt]{article}
\usepackage{times}
\usepackage{amsmath,amssymb,amsfonts,amsthm}
\usepackage{braket}
\usepackage{slashed}
\usepackage{epsfig}
\def\be{\begin{equation}}
\def\ee{\end{equation}}

\newcommand{\dd}{{\rm d}}
\newcommand{\etal}{{\it et al.}}

\newcommand{\super}[1]{$^{\ #1}$}
\newcommand{\url}[1]{[{#1}]}
\newcommand\beq{\begin{equation}}
\newcommand\eeq{\end{equation}}
\newcommand{\SigmapiN}{\ensuremath{\Sigma_{\pi\!{\scriptscriptstyle\rm N}}}}
\newcommand{\beqar}{\begin{eqnarray}}
\newcommand{\eeqar}{\end{eqnarray}}

\begin{document}
\title{\sc Gyromagnetic Factors and Atomic Clock Constraints on the Variation of Fundamental Constants}
\author{
{\sc Feng Luo}${}^{1}$\thanks{E--mail: {\tt fluo@physics.umn.edu}},
{\sc Keith A. Olive}${}^{1,2}$,\thanks{E--mail: {\tt olive@physics.umn.edu}}
{\sc Jean-Philippe Uzan}${}^{3,4,5}$\thanks{E--mail: {\tt uzan@iap.fr}}\\
{\small\em ${}^{1}$School of Physics and Astronomy, 
University of Minnesota, Minneapolis, MN 55455, USA}\\ 
{\small\em ${}^{2}$William I. Fine Theoretical Physics Institute, 
University of Minnesota, Minneapolis, MN 55455, USA}\\ 
{\small\em ${}^{3}$ Institut d'Astrophysique de Paris, UMR-7095 du CNRS, 
                       Universit\'e Paris VI Pierre et Marie Curie}\\
{\small\em  98 bis bd Arago, 75014 Paris, France}\\
{\small\em ${}^{4}$Astrophysics, Cosmology and Gravitation Centre}\\
{\small\em Department of Mathematics and Applied Mathematics, University of Cape Town} \\
{\small\em Rondebosch 7701, South Africa} \\
{\small\em ${}^{5}$ National Institute for Theoretical Physics (NITheP), Stellenbosch 7600, South Africa}}


\maketitle
\vskip -4in
\rightline{UMN--TH--3006/11, FTPI--MINN--11/16}
\vskip 4in
\begin{abstract}
We consider the effect of the coupled variations of fundamental
constants on the nucleon magnetic moment. The nucleon $g$-factor
enters into the interpretation of the
measurements of variations in the fine-structure
constant, $\alpha$,  in both the laboratory (through atomic clock measurements)
and in astrophysical systems (e.g. through measurements of the 21 cm transitions).
A null result can be translated into a limit on
the variation of a set of fundamental constants, that is usually reduced to $\alpha$. 
However, in specific models, particularly unification
models, changes in $\alpha$ are always accompanied by corresponding changes in 
other fundamental quantities such as the QCD scale, $\Lambda_{\scriptscriptstyle\rm QCD}$.  
This work tracks the changes in the nucleon $g$-factors induced from changes in $\Lambda_{\scriptscriptstyle\rm QCD}$ and
the light quark masses.  In principle, these coupled variations can improve the bounds
on the variation of $\alpha$ by an order of magnitude from {\em existing} atomic clock 
and astrophysical measurements. Unfortunately, the calculation of  the dependence 
of $g$-factors on fundamental parameters is notoriously model-dependent. 
\end{abstract}

\section{Introduction}

Any definitive measurement of a temporal or spatial
variation in a fundamental constant, such as the fine-structure constant $\alpha$,
would signal physics beyond the standard model, and in
particular a violation of the equivalence principle which is one of the foundations of
general relativity. In many cases, such an observation
would indicate the existence of a new light (usually scalar) degree of freedom \cite{Bek}.
Indeed, there has been considerable excitement during the last decade over the possible time variations in $\alpha$ from observations of quasar absorption systems
\cite{webb,murphy3,chand,quast,murphy07,chand3}. 

In effectively all unification models of non-gravitational interactions, and certainly in models
in which one imposes gauge coupling unification at some high energy scale, a
variation in $\alpha$ is invariably accompanied by variations in other gauge couplings \cite{ekow,co}.
In particular, variations in the strong gauge coupling, $\alpha_s$,
will induce variations in the  QCD scale, $\Lambda_{\scriptscriptstyle\rm QCD}$,
as can be seen from the low energy
expression for $\Lambda_{\scriptscriptstyle\rm QCD}$ when mass thresholds are included
\begin{equation}
\Lambda_{\scriptscriptstyle\rm QCD} = \mu \left(\frac{m_{\rm c} \, m_{\rm b} \, m_{\rm t}}{\mu^3} \right)^{\frac{2}{27}} \, 
\exp\left[-\frac{2\pi}{9\alpha_s(\mu)} \right] \,,
\end{equation}
for a renormalization scale $\mu > m_{\rm t}$ up to the unification scale  \cite{ekow,co,lang},
where $m_{\rm c,b,t}$ are the masses of the charm, bottom, and top quarks.
Because fermion masses are proportional to $hv$ where $h$ is a Yukawa coupling and $v$ is the Higgs vacuum expectation value (vev),
variations in Yukawa couplings will also affect variations in $\Lambda_{\scriptscriptstyle\rm QCD}$
so that 
\begin{eqnarray}\label{DeltaLambda}
\frac{\Delta \Lambda_{\scriptscriptstyle\rm QCD}}{\Lambda_{\scriptscriptstyle\rm QCD}} &= &R \, \frac{\Delta \alpha}{\alpha}
+ \frac{2}{27} \left(3 \, \frac{\Delta v}{v} + \frac{\Delta h_{\rm c}}{h_{\rm c}} + \frac{\Delta h_{\rm b}}{h_{\rm b}}
+  \frac{\Delta h_{\rm t}}{h_{\rm t}} \right) \,.
\end{eqnarray}
Typical values for $R$ are of order 30 in many grand unified theories,
but there is considerable model-dependence in this coefficient \cite{dine}.

Furthermore, in theories in which
the electroweak scale is derived by dimensional transmutation, changes in
the Yukawa couplings (particularly the top Yukawa) lead to exponentially
large changes in the Higgs vev.
In such theories, the Higgs expectation value
is related to the Planck mass, $M_{\rm P}$, by~\cite{co}
\beq\label{rad1}
   v\sim M_{\rm P} \exp\left(- \frac{2 \pi c}{ \alpha_{\rm t}}\right) ,
\eeq
where $c$ is a constant of order 1, and $\alpha_{\rm t} = h_{\rm t}^2/4\pi$.
For $c \sim h_{\rm t} \sim 1$,
\beq\label{enhance2}
{\Delta v \over v} \sim S {\Delta h_{\rm t} \over h_{\rm t}} ,
\eeq
with $S \sim 160$, though there is considerable model-dependence in this value as well.
For example, in supersymmetric models, $S$ can be related to 
the sensitivity of the $Z$ gauge boson mass to
the top Yukawa, and may take values anywhere from about 80 to 500 \cite{eos}.
This dependence gets translated into
a variation in all low energy particle masses \cite{ds}.  

In addition, in many string
theories, all gauge and Yukawa couplings are determined by the expectation value of a dilaton 
and we might expect \cite{co}
\beq\label{h-alpha}
{\Delta h \over h} = {1 \over 2} {\Delta \alpha \over \alpha} ,
\eeq
assuming that all Yukawa couplings vary similarly, so that they all
reduce to $h$. Therefore, once we
allow $\alpha$ to vary, virtually all masses and couplings are expected
to vary as well, typically much more strongly than the variation induced
by the Coulomb interaction alone.

Irrespective of the purported observations of a time variation in $\alpha$,
many experiments and analyses have led to limits on possible variations \cite{uzan,uzan2}.
Furthermore, the use of coupled variations has led to significantly improved
constraints in a wide range of environments ranging from big bang nucleosynthesis 
\cite{co,ichikawa,wett,cnouv,df,landau,grant}, the Oklo reactor \cite{opqccv,wett2}, meteoritic data \cite{opqccv,opqmvcc,wett2}, the microwave background \cite{cmb,landau}
and stellar evolution \cite{ek}.

This article explores the possibility that the strongest existing limits on the fine-structure constant,
namely those derived from atomic clock measurements, can also be enhanced by considering 
such coupled variations.
We expect the effect of induced variations in $ \Lambda_{\scriptscriptstyle\rm QCD}$ and the light quark masses
to enter through the nucleon magnetic moment.  
 Existing experimental limits on $\alpha$ from atomic clock experiments assume constant $\mu_{\rm p,n}$.
 Indeed, limits on the
variations of quark masses in units of the QCD scale, i.e. $m_{\rm q}/\Lambda_{\scriptscriptstyle\rm QCD}$,
from atomic clock measurements  
have been derived \cite{flam,lattice result ref 0402098}. Given a (model-dependent) calculation of 
the nucleon magnetic moment (or equivalently its $g$-factor), we
can derive sharper bounds on the variation of $\alpha$ from existing data.
Unfortunately, because of the model-dependence, we find 
that while the limits are generally improved (by as much as an order of magnitude),
there is considerable uncertainty in the precise numerical limit.
As a corollary, we apply our results to astrophysical measurements such as those 
which rely on the 21 cm line which also depends on $\mu_{\rm p,n}$.

The article is organized as follows:
In section 2, we outline the procedure of obtaining limits on $\alpha$ 
from atomic clock experiments. In particular, we examine the detailed dependence on
the nuclear $g$-factors which will be subject to variation. In section 3, we
derive the dependence of the nucleon magnetic moment on $\Lambda_{\scriptscriptstyle\rm QCD}$ and the light quark masses.
Because there is no unique (or rigorous) method for calculating baryon magnetic moments,
we consider several different approaches.  The most straightforward employs the constituent
quark model.  Surprisingly, this model is quite effective in matching the observed baryon magnetic 
moments. Even within this broad approach, our result will depend on 
the calculation of the nucleon mass, as well as the calculation of the constituent quark mass; each 
carrying a significant degree of uncertainty.  We also consider an approach based 
on chiral perturbation theory, and a method based partially on 
lattice results.  In section 4, we apply these results to atomic clock measurements
and derive ``improved" limits on the variation of $\alpha$.  Finally, in section 5,
we extend these results to measurements involving the 21 cm line and summarize our results.

\section{Atomic clock constraints}\label{Section2}

\subsection{From frequency shifts to constants}\label{Section2A}

The comparison of atomic clocks provides a constraint on the relative shift
of the frequencies of the two clocks as a function of time, on
time scales of the order of a couple of years. This observation (or lack thereof)
can be translated into a constraint on the time variation
of a fundamental constant. Using QED, the frequency of the
atomic transitions can be expressed (see e.g. \cite{clock-bize05}) in terms of the fine
structure constant $\alpha$, the electron-to-proton mass ratio,
$\mu\equiv m_{\rm e}/m_{\rm p}$ and the gyromagnetic factor $g_i=2 \mu_i/\mu_{\mathrm{N}}$,
where $\mu_i$ is the nuclear magnetic moment, and $\mu_{\mathrm{N}} = \frac{e}{2 m_{\rm p}}$ is the nuclear magneton.

The hyperfine frequency in a given electronic state of an alkali-like
atom is given by
\begin{equation}\label{e.hf}
 \nu_{\mathrm{hfs}} \simeq R_\infty c \times A_{\mathrm{hfs}} \times g_i
 \times \alpha^2\times\mu \times F_{\mathrm{hfs}}(\alpha),
\end{equation}
where $R_\infty$ the Rydberg constant, $A_{\mathrm{hfs}}$
is a numerical factor depending on the atomic species and 
$F_{\mathrm{hfs}}(\alpha)$ is a factor taking into account relativistic
corrections (including the Casimir contribution) which depends on the atom. 
We omitted the effect of the finite nuclear radius on hyperfine frequency in Eq.~(\ref{e.hf}), since the effect of varying the nuclear radius is shown to be smaller~\cite{nuclear radius effect, BFK2011} than the effects of varying other parameters which we consider in this work.    
Similarly, the frequency of an
electronic transition is well-approximated by
\begin{equation}\label{e.elec}
 \nu_{\text{elec}} \simeq R_\infty c \times A_{\text{elec}} \times F_{\text{elec}}(Z,\alpha),
\end{equation}
where, as above, $A_{\text{elec}}$ is a numerical factor depending on
each particular atom and $F_{\text{elec}}$ is the function accounting
for relativistic effects, spin-orbit couplings and many-body
effects. Even though an electronic transition should also include
a contribution from the hyperfine interaction, it is generally
only a small fraction of the transition energy and thus should not
carry any significant sensitivity to a variation of the
fundamental constants.

Relativistic corrections are important~\cite{clock-prestage} and are
computed by means of relativistic $N$-body 
calculations~\cite{kappa-dzuba2, kappa-dzuba3, kappa-dzuba, kappa-flamb}. 
These can be characterized by introducing the
sensitivity of the relativistic factors to a variation of $\alpha$ defined by
\begin{equation}
 \kappa_\alpha = \frac{\delta\ln F}{\delta\ln\alpha}.
\end{equation}
The values of these coefficients for the transitions that we shall consider below are
summarized in Table~\ref{tab0}.

\begin{table}[t]
\caption{Sensitivity of various transitions on a variation of
the fine structure constant. From Refs.~~\cite{kappa-dzuba2, kappa-dzuba3, kappa-dzuba, kappa-flamb}.}
\label{tab0}
\centering
{\small
\begin{tabular}{|p{4.0cm}cc|}
\hline
 Atom  & Transition & Sensitivity $\kappa_\alpha$ \\
 \hline\hline
 \super{1}H            & $1s-2s$                  &  0.00 \\
 \super{87}Rb           &       hf                 & 0.34   \\
 \super{133}Cs          & ${}^2S_{1/2}(F=2)-(F=3)$  & 0.83   \\
 \super{171}Yb\super{+} & ${}^2S_{1/2}-{}^2D_{3/2}$  & 0.9    \\
 \super{199}Hg\super{+} & ${}^2S_{1/2}-{}^2D_{5/2}$  & --3.2   \\
 \super{87}Sr           & ${}^1S_0-{}^3P_0$        & 0.06       \\
 \super{27}Al\super{+}  & ${}^1S_0-{}^3P_0$        & 0.008    \\
 \hline\hline
\end{tabular}
}
\end{table}

\subsection{Experimental constraints}\label{Section2B}

Over the past several years, many comparisons of atomic clocks have been performed.
We consider only the latest result of each type of comparison for our analysis.
\begin{itemize}
\item \emph{Rubidium}: The comparison of the hyperfine frequencies of rubidium and caesium in their electronic
ground state between 1998 and 2004~\cite{clock-bize05} yields
\begin{equation} \label{rbcs}
  \frac{\dd}{\dd t}\ln\left(\frac{\nu_{\rm Cs}}{\nu_{\rm Rb}}\right)
             = (0.5\pm5.3)\times10^{-16}\,\rm{yr}^{-1}.
\end{equation}
From Eq.~(\ref{e.hf}), and using the values of the sensitivities $\kappa_\alpha$, we deduce that this comparison constrains
\begin{equation}  \label{nucsrb}
   \frac{\nu_{\rm Cs}}{\nu_{\rm Rb}}\propto\frac{g_{\rm Cs}}{g_{\rm Rb}}\,\alpha^{0.49}.
\end{equation}

\item\emph{Atomic hydrogen}: The $1s-2s$ transition in atomic hydrogen was compared to the ground state hyperfine
splitting of caesium~\cite{clock-fisher04} in 1999 and 2003, setting an upper limit on the variation of $\nu_{\rm H}$ of $(-29\pm57)$~Hz within 44~months. This can be translated in a relative drift
\begin{equation}\label{clock-H}
  \frac{\dd}{\dd t}\ln\left(\frac{\nu_{\rm Cs}}{\nu_{\rm H}}\right)
    = (32\pm63)\times10^{-16}\,\rm{yr}^{-1}.
\end{equation}
Since the relativistic correction for the atomic hydrogen transition nearly vanishes, we have $\nu_{\rm H}\sim R_\infty$
so that
 \begin{equation}\label{nuhcs}
  \frac{\nu_{\rm Cs}}{\nu_{\rm H}}\propto g_{\rm Cs}\,\mu\,\alpha^{2.83}.
\end{equation}

\item \emph{Mercury}: The $^{199}$Hg$^+$ ${}^2S_{1/2}-{}^2D_{5/2}$ optical transition has a high sensitivity to $\alpha$ (see Table~\ref{tab0}) so that it is well suited to test its variation. The frequency of the $^{199}$Hg$^+$ electric quadrupole transition at 282~nm was thus compared to the ground state hyperfine transition of caesium first during a two year period~\cite{clock-bize03} and then over a 6 year period~\cite{clock-fortier07} to get
\begin{equation} \label{hgcs}
   \frac{\dd}{\dd t}\ln\left(\frac{\nu_{\rm Cs}}{\nu_{\rm Hg}}\right)=
    (-3.7\pm3.9)\times10^{-16}\,\rm{yr}^{-1}.
\end{equation}
While $\nu_{\rm Cs}$ is still given by Eq.~(\ref{e.hf}), $\nu_{\rm Hg}$ is given by Eq.~(\ref{e.elec}). Using the sensitivities of Table~\ref{tab0}, we conclude that this comparison test the stability of
\begin{equation}\label{nuhgcs}
 \frac{\nu_{\rm Cs}}{\nu_{\rm Hg}}\propto g_{\rm Cs}\,\mu\,\alpha^{6.03}.
\end{equation}

\item \emph{Ytterbium}: The ${}^2S_{1/2}-{}^2D_{3/2}$ electric quadrupole transition at 688~THz of $^{171}$Yb$^+$ was compared to  the ground state hyperfine transition of  caesium. The constraint of \cite{clock-peik04} was updated,
  after a comparison over a six year period, which leads to~\cite{clock-peik06}
\begin{equation}\label{ybcs}
   \frac{\dd}{\dd t}\ln\left(\frac{\nu_{\rm Cs}}{\nu_{\rm Yb}}\right)=
    (0.78\pm1.40)\times10^{-15}\,\rm{yr}^{-1}.
\end{equation}
 This tests the stability of
\begin{equation} \label{nuybcs}
 \frac{\nu_{\rm Cs}}{\nu_{\rm Yb}}\propto g_{\rm Cs}\,\mu\,\alpha^{1.93}.
\end{equation}

\item \emph{Strontium}:  The comparison of the ${}^1S_0-{}^3P_0$ transition in neutral ${}^{87}$Sr with a caesium clock was performed in three independent laboratories. The combination of these three experiments~\cite{clock-blatt} leads to the constraint
\begin{equation}  \label{srcs}
   \frac{\dd}{\dd t}\ln\left(\frac{\nu_{\rm Cs}}{\nu_{\rm Sr}}\right)=
    (1.0\pm1.8)\times10^{-15}\,\rm{yr}^{-1}.
\end{equation}
Similarly, this tests the stability of
\begin{equation}\label{nusrcs}
 \frac{\nu_{\rm Cs}}{\nu_{\rm Sr}}\propto g_{\rm Cs}\,\mu\,\alpha^{2.77}.
\end{equation}

\item \emph{Atomic dyprosium}: The electric dipole (E1) transition between two nearly degenerate opposite-parity states in atomic dyprosium should be highly sensitive to the variation of $\alpha$~\cite{kappa-dzuba3,kappa-dzuba,Dy-df,nguyen04}. The frequencies of two isotopes of dyprosium were monitored over a 8 months period~\cite{clock-cingoz} showing that the frequency variation of the 3.1-MHz transition in $^{163}$Dy and the 235-MHz transition in $^{162}$Dy are 9.0$\pm$6.7 Hz/yr and -0.6$\pm$6.5 Hz/yr, respectively. This provides the constraint
\begin{equation}\label{clock-dypro}
  \frac{\dot\alpha}{\alpha} =(-2.7\pm2.6)\times 10^{-15}\,\rm{yr}^{-1},
\end{equation}
at 1$\sigma$ level, without any assumptions on the constancy of other fundamental constants.  

\item \emph{Aluminium and mercury single-ion optical clocks}: The comparison of the ${}^1S_{0}- {}^3P_{0}$ transition in ${}^{27}$Al$^+$ and ${}^2S_{1/2}-{}^2D_{5/2}$ in ${}^{199}$Hg$^+$ over a year allowed one to set the constraint~\cite{clock-rosen08}
\begin{equation}
  \frac{\dd}{\dd t}\ln\left(\frac{\nu_{\rm Al}}{\nu_{\rm
   Hg}}\right)= (-5.3\pm7.9)\times10^{-17}\,\rm{yr}^{-1}.
\end{equation}
  Proceeding as previously, this tests the stability of
\begin{equation}\label{e3.208}
 \frac{\nu_{\rm Al}}{\nu_{\rm Hg}}\propto \alpha^{3.208},
\end{equation}
which, using Eq.~(\ref{e3.208}) directly sets the constraint
\begin{equation}\label{clock-bound1}
  \frac{\dot\alpha}{\alpha} =(-1.65\pm2.46)\times10^{-17}\,\rm{yr}^{-1},
\end{equation}
since it depends only on $\alpha$.
\end{itemize}

Experiments with diatomic molecules, as first
pointed out by Thomson~\cite{mu-theorie}
provide a test of the variation of $\mu$. The energy difference
between two adjacent rotational levels in a diatomic molecule is
inversely proportional to $M r^{-2}$, $r$ being the bond length
and $M$ the reduced mass, and the vibrational transition of the
same molecule has, in first approximation, a $\sqrt{M}$
dependence. For molecular hydrogen $M=m_{\rm p}/2$ so that the
comparison of an observed vibro-rotational spectrum with a
laboratory spectrum gives an information on the variation
of $m_{\rm p}$ and $m_{\rm n}$. Comparing pure rotational
transitions with electronic transitions gives a measurement of
$\mu$. It follows that the frequency of vibro-rotation transitions
is, in the Born-Oppenheimer approximation, of the form
\begin{equation}\label{mu1}
\nu\simeq E_I\left(c_{_{\rm elec}} +c_{_{\rm vib}}\sqrt{\mu}
+c_{_{\rm rot}}\mu\right),
\end{equation}
where $c_{_{\rm elec}}$, $c_{_{\rm vib}}$ and $c_{_{\rm rot}}$ are
some numerical coefficients.

The comparison of the vibro-rotational transition in the molecule
SF6 was compared to a caesium clock over a two-year period,
leading to the constraint~\cite{clock-mu}
\begin{equation}  \label{sf6cs}
   \frac{\dd}{\dd t}\ln\left(\frac{\nu_{\rm Cs}}{\nu_{\rm SF6}}\right)=
    (-1.9\pm0.12\pm2.7)\times10^{-14}\,\rm{yr}^{-1},
\end{equation}
where the second error takes into account uncontrolled
systematics. Now, using Table~\ref{tab0} again and Eq.~(\ref{e.hf}) for Cs, we deduce that
for a vibrational transition,
\begin{equation} \label{nusf6cs}
 \frac{\nu_{\rm Cs}}{\nu_{\rm SF6}}\propto g_{\rm Cs}\, \sqrt{\mu}\,\alpha^{2.83}.
\end{equation}

\subsection{Nuclear $g$-factors}\label{Section2C}

All the constraints involve only 4 quantities, $\mu$, $\alpha$ and the two gyromagnetic factors
$g_{\rm Cs}$ and $g_{\rm Rb}$.
It follows that we need to relate the nuclear $g$-factors that appeared in the constraints
of the previous subsection, with the proton and neutron $g$-factors that will
be calculated in Section~\ref{Section3}.

An approximate calculation of the nuclear magnetic moment is possible
in the shell model and is relatively simple for even-odd (or odd-even) nuclei
where the nuclear magnetic moment is determined by the unpaired nucleon.
For a single nucleon, in a particular $(l,j)$ state within the nucleus, 
we can write
\begin{eqnarray}\label{nucg}
&& g =\left\lbrace
\begin{array}{l}
2 l g_l + g_s \\
\frac{j}{j+1} [2(l+1) g_l - g_s]
\end{array}
\right.
\qquad \hbox{for}\qquad
\left\lbrace
\begin{array}{l}
j = l + \frac12 \\
j = l - \frac12
\end{array}
\right.
\end{eqnarray}
where $g_l = 1 (0) $ and $g_s = g_{\rm p} (g_{\rm n})$ for a valence proton (neutron).  

From the previous discussion, the only $g$-factors that are needed are those for \super{87}Rb and \super{133}Cs. For both isotopes, we have an unpaired valence proton.  For \super{87}Rb, the ground state is in a $p_{3/2}$ state so that $l = 1$ and $j = \frac32$, while for \super{133}Cs, 
the ground state is in  a $g_{7/2}$ state corresponding to $l= 4$ and $j=\frac72$. 
Using Eq.~(\ref{nucg}), the nuclear $g$-factor can easily be expressed in terms of $g_{\rm p}$ alone. Using $g_{\rm p} = 5.586$, we find $g = 7.586$ for \super{87}Rb and $g = 3.433$ for \super{133}Cs, while the experimental values are $g = 5.502$ for \super{87}Rb and $g = 5.164$ for \super{133}Cs. 

The differences between the shell model predicted $g$-factors and the experimental values can be attributed to the effects of the polarization of the non-valence nucleons and spin-spin interaction~\cite{flam, BFK2011}. Taking these effects into account, the refined formula relevant for our discussion of \super{87}Rb and \super{133}Cs is
\be
\label{refined g-factor formula}
g =  2 \left[ g_{\rm n} \, b \, \langle s_z \rangle^o + (g_{\rm p} - 1) (1-b) \langle s_z \rangle^o +  j \right]  , 
\ee
where $g_{\rm n} = -3.826$, $\langle s_z \rangle^o$ is the spin expectation value of the single valence proton in the shell model and it is one half of the coefficient of $g_s$ in Eq.~(\ref{nucg}), and $b$ is determined by the spin-spin interaction and it appears in the expressions for the spin expectation value of the valence proton $\langle s_{z_{\rm p}} \rangle = (1-b) \langle s_z \rangle^o$ and non-valence neutrons $\langle s_{z_{\rm n}} \rangle = b \langle s_z \rangle^o$. Following the preferred method in~\cite{flam, BFK2011}, it is found 
\be
\label{szn formula}
\langle s_{z_{\rm n}} \rangle = \frac{\frac{g}{2} - j - (g_{\rm p} -1) \langle s_z \rangle^o}{g_{\rm n} + 1 - g_{\rm p}} \, , 
\ee
and
\be
\label{szp formula}
 \langle s_{z_{\rm p}} \rangle =  \langle s_z \rangle^o -  \langle s_{z_{\rm n}} \rangle . 
\ee
Therefore, the variation of the $g$-factor can be written as
\be
\label{preferred method formula}
\frac{\delta g}{g} = \frac{\delta g_{\rm p}}{g_{\rm p}} \frac{2 g_{\rm p} \langle s_{z_{\rm p}} \rangle}{g} +  \frac{\delta g_{\rm n}}{g_{\rm n}} \frac{2 g_{\rm n} \langle s_{z_{\rm n}} \rangle}{g} + \frac{\delta b}{b} \frac{2 (g_{\rm n} - g_{\rm p} + 1)  \langle s_{z_{\rm n}} \rangle}{g} \, . 
\ee
From Eq.~(\ref{szn formula}), (\ref{szp formula}) and (\ref{preferred method formula}), we find, by using the experimental $g$-factors, 
\begin{eqnarray}
\label{new eq27}
\frac{\delta g_{\rm Rb}}{g_{\rm Rb}} &=& 0.764 \frac{\delta g_{\rm p}}{g_{\rm p}} - 0.172 \frac{\delta g_{\rm n}}{g_{\rm n}} - 0.379 \frac{\delta b}{b} \, , \\
\label{new eq28}
\frac{\delta g_{\rm Cs}}{g_{\rm Cs}} &=& -0.619 \frac{\delta g_{\rm p}}{g_{\rm p}} + 0.152 \frac{\delta g_{\rm n}}{g_{\rm n}} + 0.335 \frac{\delta b}{b} \, .
\end{eqnarray}

\subsection{Summary of the constraints}\label{Section2D}

Given the discussion in the two previous subsections, and in particular Eqs.~(\ref{new eq27})
and~(\ref{new eq28}), the atomic clock experiments give constraints on
the set $\lbrace g_{\rm p}, g_{\rm n}, b, \mu, \alpha\rbrace$ and thus variations in the relative frequency shift $\nu_{AB}=\nu_A/\nu_B$
are given by
\begin{equation}\label{decomposition1}
 \frac{\delta\nu_{AB}}{\nu_{AB}} = \lambda_{g_{\rm p}}  \frac{\delta g_{\rm p}}{g_{\rm p}}
 + \lambda_{g_{\rm n}}  \frac{\delta g_{\rm n}}{g_{\rm n}}
 + \lambda_b  \frac{\delta b}{b}
  + \lambda_{\mu}  \frac{\delta\mu}{\mu} +\lambda_{\alpha}  \frac{\delta \alpha}{\alpha},
\end{equation}
or equivalently
\begin{equation}\label{decomposition1t}
 \frac{\dot\nu_{AB}}{\nu_{AB}} = \lambda_{g_{\rm p}}  \frac{\dot g_{\rm p}}{g_{\rm p}}
 +\lambda_{g_{\rm n}}  \frac{\dot g_{\rm n}}{g_{\rm n}}
 +\lambda_b  \frac{\dot b}{b}
  + \lambda_{\mu}  \frac{\dot\mu}{\mu} +\lambda_{\alpha}  \frac{\dot \alpha}{\alpha},
\end{equation}
with the coefficients $\lbrace\lambda_{g_{\rm p}}, \lambda_{g_{\rm n}}, \lambda_b, \lambda_{\mu},  \lambda_{\alpha} \rbrace$
summarized in Table~\ref{tab:coef1}.

For the sake of comparison, the shell model gives
\be
\label{shell model prediction for Rb}
\frac{\delta g_{\rm Rb}}{ g_{\rm Rb}} \simeq 0.736\frac{\delta g_{\rm
p}}{g_{\rm p}}
\ee
and
\be
\label{shell model prediction for Cs}
 \frac{\delta g_{\rm Cs}}{ g_{\rm Cs}} \simeq -1.266  \frac{\delta g_{\rm
p}}{g_{\rm p}}.
\ee
The main difference arises from the dependence in $g_{\rm n}$ and $b$ but
the order
of magnitude is similar.

\begin{table}[htb!]
\caption{Summary of the constraints of the atomic clock experiments and values of the coefficients
$\lbrace\lambda_{g_{\rm p}},  \lambda_{g_{\rm n}}, \lambda_b, \lambda_{\mu},  \lambda_{\alpha} \rbrace$ entering
the decomposition~(\ref{decomposition1t}).}
\label{tab:coef1}
\begin{center}
\begin{tabular}{|l|c|ccccc|c|}
\hline
 Clocks   & $\nu_{AB}$ & $\lambda_{g_{\rm p}}$ &  $\lambda_{g_{\rm n}}$ &  $\lambda_b$ & $\lambda_{\mu}$ & $\lambda_{\alpha}$ 
       & $\dot\nu_{AB}/\nu_{AB}$ (yr$^{-1}$) \\
\hline\hline
 Cs - Rb & $\frac{g_{\rm Cs}}{g_{\rm Rb}}\,\alpha^{0.49}$    & $-1.383$   &  $0.325$ & $0.714$ &  0    &  $0.49$  &  $(0.5\pm5.3)\times10^{-16}$ \\
 Cs - H &  $g_{\rm Cs}\,\mu\,\alpha^{2.83}$   &  $-0.619$  & $0.152$ & $0.335$ &   1   & $2.83$   & $(32\pm63)\times10^{-16}$ \\
 Cs - $^{199}$Hg$^+$  & $g_{\rm Cs}\,\mu\,\alpha^{6.03}$    & $-0.619$  & $0.152$ & $0.335$   &   1    & $6.03$   &  $(-3.7\pm3.9)\times10^{-16}$ \\
 Cs - $^{171}$Yb$^+$ & $g_{\rm Cs}\,\mu\,\alpha^{1.93}$    & $-0.619$  & $0.152$ & $0.335$  &    1    & $1.93$   &  $ (0.78\pm1.40)\times10^{-15}$ \\ 
 Cs - Sr &  $g_{\rm Cs}\,\mu\,\alpha^{2.77}$   &  $-0.619$  & $0.152$ & $0.335$ &      $1$  &  $2.77$  & $(1.0\pm1.8)\times10^{-15}$ \\
 Cs - SF$_6$  &  $g_{\rm Cs}\sqrt{\mu}\alpha^{2.83}$   & $-0.619$  & $0.152$ & $0.335$ &   0.5     &  2.83  &  $ (-1.9\pm0.12\pm2.7)\times10^{-14}$ \\
 Dy &  $\alpha$   & 0   &  0  & 0 & 0   & 1   & $(-2.7\pm2.6)\times 10^{-15}$  \\
   $^{199}$Hg$^+$ - $^{27}$Al$^+$ & $\alpha^{-3.208}$    &  0 & 0 & 0 &  0      &  $-3.208$  & $(5.3\pm7.9)\times10^{-17}$ \\
  \hline
\end{tabular}
\end{center}
\end{table}


\section{Nucleon magnetic moments, current quark masses and $\Lambda_{\scriptscriptstyle\rm QCD}$}\label{Section3}

In this section, we will review several approaches in the literature in calculating the nucleon magnetic moments, including the non-relativistic constituent quark model (NQM), chiral perturbation theory ($\chi$PT), and a method combining the results of $\chi$PT and lattice QCD. We will try to extract the dependence of the nucleon magnetic moments on the current quark masses and $\Lambda_{\scriptscriptstyle\rm QCD}$ from the expressions given by each of these approaches.

\subsection{The non-relativistic constituent quark model approach}

The NQM, which approximates hadrons as bound states of their constituent quarks gives a good approximation to the measured baryon magnetic moments~\cite{constituent_bmm}. In this model, the baryon magnetic moments are expressed in terms of the Dirac magnetic moments of their constituent quarks, with the coefficients given by the baryon spin/flavor wave functions. For the proton and neutron, the magnetic moments are
\be
\label{constituent_starting eq}
\mu_{\rm p} = \frac{4}{3} \mu_{\rm u} - \frac{1}{3} \mu_{\rm d}  \;\;\;\;\;\;\; \text{and} \;\;\;\;\;\;\; 
\mu_{\rm n} = \frac{4}{3} \mu_{\rm d} - \frac{1}{3} \mu_{\rm u} \, ,
\ee
where $\mu_{\rm u} = \frac{2}{3}\frac{e}{2M_{\rm u}}$ and $\mu_{\rm d} = -\frac{1}{3}\frac{e}{2M_{\rm d}}$. Here,  $M_{\rm u}$ and $M_{\rm d}$ are the {\it constituent} $u$ and $d$ quark masses, respectively, with their values around a third of the nucleon mass, to be compared with the much smaller $u$ and $d$ {\it current} quark masses, $m_{\rm u}$ and $m_{\rm d}$, which are several MeV. For the three light flavors ($u$, $d$ and $s$), the main part of their constituent quark masses have a strong interaction origin, with the dynamics of the virtual gluons and quark-antiquark sea being responsible for the large masses~\cite{Fritzsch:2009xe}, while the current quark masses which contribute only a small portion of their corresponding constituent quark masses are of pure electroweak origin.

From Eq.~(\ref{constituent_starting eq}), the nucleon magnetic moment in units of the nuclear magneton $\mu_{\mathrm{N}} = \frac{e}{2 m_{\rm p}}$,  that is, the $g$-factor of the nucleon, can be written as 
\be
\label{constituent_g factor eq}
g_{\scriptscriptstyle\rm NQM} =2 \left(c_{\rm u} \frac{m_{\rm p}}{M_{\rm u}} + c_{\rm d} \frac{m_{\rm p}}{M_{\rm d}}\right),
\ee
where $c_{\rm u} = 8/9$ and $c_{\rm d} = 1/9$ for the proton, and $c_{\rm u} = -2/9$ and $c_{\rm d} = -4/9$ for the neutron. In the study of hadron properties, the constituent quark masses are usually taken as fitting parameters, with $M_{\rm u} = M_{\rm d}$ often assumed~\cite{constituent_bmm}, since isospin is a good approximate symmetry. We will assume this relation in the following calculations to simplify the algebra, but we emphasize that $\delta M_{\rm u}$ may not necessarily be equal to $\delta M_{\rm d}$. By differentiating Eq.~(\ref{constituent_g factor eq}), we obtain a general expression for the variation of the $g$-factor
\be
\label{constituent_vary starting eq}
\frac{\delta g_{\scriptscriptstyle\rm NQM}}{g_{\scriptscriptstyle\rm NQM}} = \frac{\delta m_{\rm p}}{m_{\rm p}} - \left(\frac{c_{\rm u}}{c_{\rm u} + c_{\rm d}} \frac{\delta M_{\rm u}}{M_{\rm u}} + \frac{c_{\rm d}}{c_{\rm u} + c_{\rm d}} \frac{\delta M_{\rm d}}{M_{\rm d}} \right)  .
\ee

The proton mass, $m_{\rm p}$, and $M_{\rm u,d}$ are functions of the fundamental constants, and they can be formally written as  $m_{\rm p} = m_{\rm p} (v_1, v_2, \cdots, v_n)$ and $M_{\rm u, d} = M_{\rm u, d} (v_1, v_2, \cdots, v_n)$, where the $v_i$'s are fundamental constants including $m_{\rm u}$, $m_{\rm d}$, $m_{\rm s}$, $\Lambda_{\scriptscriptstyle\rm QCD}$, etc.. Therefore, Eq.~(\ref{constituent_vary starting eq}) becomes
\begin{eqnarray}
\label{constituent_general vary function eq}
\frac{\delta g_{\scriptscriptstyle\rm NQM}}{g_{\scriptscriptstyle\rm NQM}} & = & \sum_{i=1}^n \frac{\delta v_i}{v_i} \left[\frac{v_i}{m_{\rm p}}\frac{\partial m_{\rm p}}{\partial v_i} - \left(\frac{c_{\rm u}}{c_{\rm u} + c_{\rm d}} \frac{v_i}{M_{\rm u}} \frac{\partial M_{\rm u}}{\partial v_i} + \frac{c_{\rm d}}{c_{\rm u} + c_{\rm d}} \frac{v_i}{M_{\rm d}} \frac{\partial M_{\rm d}}{\partial v_i} \right) \right]  \nonumber \\
&& \equiv \sum_{i=1}^n \frac{\delta v_i}{v_i} \kappa_i.\label{defkappa}
\end{eqnarray}
This is our key equation in studying the dependence of the $g$-factors on fundamental constants in the NQM approach, and the problem amounts to finding the expressions for  $m_{\rm p} (v_1, \cdots, v_n)$ and 
$M_{\rm u,d} (v_1, \cdots, v_n)$. 

\subsubsection{The current quark mass and $\Lambda_{\scriptscriptstyle\rm QCD}$ dependence of $m_{\rm p}$}
\label{sec: fTqg}

To get the coefficients of $\frac{\delta m_{\rm p}}{m_{\rm p}}$, that is, the first term in the square bracket of Eq.~(\ref{constituent_general vary function eq}), we follow the procedure of~\cite{HY Cheng, sigma term ref1, sigma term ref2}, by defining $B_{\rm q}$ $(q=u,d,s)$ and the $\pi$-nucleon sigma term, $\SigmapiN$, in terms of proton matrix elements,
\beqar
\label{Bq definition}
m_{\rm q} B_{\rm q} &\equiv& \langle p| m_{\rm q} \bar{q} q | p \rangle = m_{\rm q} \frac{\partial m_{\rm p}}{\partial m_{\rm q}},
\\
\label{Sigma definition}
\SigmapiN &\equiv& \langle p|  \hat{m} (\bar{u}u + \bar{d}d) | p \rangle = \hat{m} \frac{\partial m_{\rm p}}{\partial \hat{m}} \, ,
\eeqar
where $\hat{m} \equiv \frac{1}{2} (m_{\rm u} + m_{\rm d})$. The latter equalities of the above two equations come from the Hellmann-Feynman theorem~\cite{HF theorem ref} as noted by Gasser~\cite{FH Gasser ref}. 

By using the strangeness fraction of the proton, 
\be
\label{strangeness fraction definition}
y \equiv{2B_{\rm s} \over B_{\rm d} + B_{\rm u}} = 1-{\sigma_0 \over \SigmapiN} \, ,
\ee
where $\sigma_0$ is the shift in the nucleon mass due to nonzero quark masses, and a relation from the energy-momentum tensor trace anomaly~\cite{trace anomaly ref}  for the baryon-octet members~\cite{SVZ, HY Cheng, Li and Cheng},
\be
\label{trace anomaly octet eq}
z \equiv {B_{\rm u} - B_{\rm s} \over B_{\rm d} - B_{\rm s}} =
{m_{\Xi^0} + m_{\Xi^-} -m_{\rm p} -m_{\rm n} \over 
m_{\Sigma^+} + m_{\Sigma^-} -m_{\rm p} -m_{\rm n}} \approx 1.49,
\ee
we can derive from Eqs.~(\ref{Bq definition}) and (\ref{Sigma definition}) the current quark masses dependence of $m_{\rm p}$, denoted as $f_{T_{\rm q}}$'s, as
\beqar
\label{fTq's expression eq}
f_{T_{\rm u}} \equiv \frac{m_{\rm u} B_{\rm u}}{m_{\rm p}} &=& \frac{2 \SigmapiN}{m_{\rm p} \left(1+ \frac{m_{\rm d}}{m_{\rm u}}\right) \left(1+\frac{B_{\rm d}}{B_{\rm u}}\right)} \, , \nonumber \\
f_{T_{\rm d}} \equiv \frac{m_{\rm d} B_{\rm d}}{m_{\rm p}} &=& \frac{2 \SigmapiN}{m_{\rm p} \left(1+ \frac{m_{\rm u}}{m_{\rm d}}\right) \left(1+\frac{B_{\rm u}}{B_{\rm d}}\right)} \, ,  \\
f_{T_{\rm s}}  \equiv \frac{m_{\rm s} B_{\rm s}}{m_{\rm p}} &=& \frac{\left(\frac{m_{\rm s}}{m_{\rm d}}\right) \SigmapiN \, y}{m_{\rm p} \left(1 + \frac{m_{\rm u}}{m_{\rm d}}\right)} \, , \nonumber
\eeqar
where
\be
{B_{\rm d} \over B_{\rm u}} = {2 + y(z - 1)  \over 2  z - y(z - 1) } \,.
\ee

Motivated by the trace anomaly expression for $m_{\rm p}$, 
\be
\label{trace anomaly eq proton}
m_{\rm p} = m_{\rm u} B_{\rm u} + m_{\rm d} B_{\rm d} + m_{\rm s}  B_{\rm s} + \hbox{gluon \; term} \, ,
\ee
we will write the remaining fundamental constants dependence of $m_{\rm p}$ as 
\be
\label{fTg expression eq}
f_{T_{\rm g}} \equiv \frac{\Lambda_{\scriptscriptstyle\rm QCD}}{m_{\rm p}} \frac{\partial m_{\rm p}}{\partial \Lambda_{\scriptscriptstyle\rm QCD}} = 1-\sum_{q=u,d,s} f_{T_{\rm q}} \, ,
\ee
which is the coefficient of $\delta \Lambda_{\scriptscriptstyle\rm QCD} / \Lambda_{\scriptscriptstyle\rm QCD}$ in $\delta m_{\rm p} / m_{\rm p}$.  
The argument behind Eq.~(\ref{fTg expression eq}) is the following: the $gluon \; term$ has its origin in the strong interaction, and $\Lambda_{\scriptscriptstyle\rm QCD}$, which is approximately the scale at which the strong interaction running coupling constant diverges, is the only mass parameter of the strong interaction in the chiral limit $m_{\rm u} = m_{\rm d} = m_{\rm s} = 0$, and therefore in this limit all of the other finite mass scales of the strong interaction phenomena, including pion decay constant, the spontaneous chiral symmetry breaking scale, etc., are related to $\Lambda_{\scriptscriptstyle\rm QCD}$ by some pure number of order one~\cite{Gasser and Leutwyler quark masses}. Note that the heavy quark $(c, b, t)$ masses do not explicitly appear in Eq.~(\ref{trace anomaly eq proton}) as discussed in~\cite{SVZ}. Then as the only other variable besides the light current quark masses, we get Eq.~(\ref{fTg expression eq}) for the $\Lambda_{\scriptscriptstyle\rm QCD}$ dependence in $m_{\rm p}$. 
We note that the $f_{T_{\rm q}}$'s and the $f_{T_{\rm g}}$ are also needed in the next section when we vary the electron-to-proton mass ratio, $\mu \equiv m_{\rm e} / m_{\rm p}$. 

In calculating the $f_{T_{\rm q}}$'s and the $f_{T_{\rm g}}$, we take the central values given in~\cite{Leutwyler 96} for the current quark mass ratios, $\frac{m_{\rm u}}{m_{\rm d}} = 0.553$ and $\frac{m_{\rm s}}{m_{\rm d}} = 18.9$, the central value of the $\pi$-nucleon sigma term suggested in~\cite{sigma term ref2}, $\SigmapiN = 64 \, \text{MeV}$, and we take $\sigma_0 = 36 \, \text{MeV}$~\cite{sigma_0 ref,morespn} and $m_{\rm p} = 938.3 \, \text{MeV}$. The results are 
\be
\label{fTqg result}
f_{T_{\rm u}} = 0.027, \quad f_{T_{\rm d}} = 0.039, \quad f_{T_{\rm s}} = 0.363, \quad f_{T_{\rm g}} = 0.571. 
\ee

In the isospin-symmetric limit such that $m_{\rm u} = m_{\rm d} = \hat{m}$, which will be needed in subsections 3.2 and 3.3, Eqs.~(\ref{fTq's expression eq}) and ({\ref{fTg expression eq}}) take simpler forms, 
\be
\label{fTqg isospin eq}
f_{T \hat{m}} = \frac{\SigmapiN}{m_{\rm p}} \, , \quad f_{T_{\rm s}} = \frac{\frac{m_{\rm s}}{\hat{m}} \SigmapiN y}{2 m_{\rm p}} \, , \quad f_{T_{\rm g}} = 1 - f_{T \hat{m}} - f_{T_{\rm s}} \, .
\ee
In calculating the values for this isospin-symmetric limit case, we take $\frac{m_{\rm s}}{\hat{m}} = 25$~\cite{PDG ref}, and the results are
\be
\label{fTqg isospin eq 2}
f_{T\hat{m}} = 0.068, \quad f_{T_{\rm s}} = 0.373, \quad f_{T_{\rm g}} = 0.559. 
\ee

\subsubsection{Expressions for $M_{\rm u,d}$ without an explicit quark sea}
\label{sec: intuitive NQM}

To get the coefficients of $\frac{\delta M_{\rm u}}{M_{\rm u}}$ and $\frac{\delta M_{\rm d}}{M_{\rm d}}$, we need to model the constituent quark masses. 
Intuitively, $M_{\rm u,d}$ can be written as 
\be
\label{cons intuitive eq}
M_{\rm q} = m_{\rm q} + a_{\rm q,int} \Lambda_{\scriptscriptstyle\rm QCD} \;\;\;\;\; (q=u,d)  , \qquad (A)
\ee
where $a_{\rm q,int}$'s are pure dimensionless numbers. The argument behind this form is the following: if the strong interaction were switched off, the constituent quark mass would be identical to its corresponding valence current quark mass which is obtained from the electroweak symmetry breaking. On the other hand, in the chiral limit, $m_{\rm u} = m_{\rm d} = m_{\rm s} = 0$, the strong interaction is responsible for the entire constituent quark mass. The above intuitive expression for the constituent quark masses does not {\it explicitly} take into account the sea quark contribution, which if included will depend on the current quark masses, similar to the terms $m_{\rm q} B_{\rm q}$ in the proton mass trace anomaly formula Eq.~(\ref{trace anomaly eq proton})~\cite{HY Cheng}. However, one could argue that the sea quark contribution is already included {\it implicitly} in the second term of Eq.~(\ref{cons intuitive eq}) together with the virtual gluons contribution, since the dynamics of the quark sea and virtual gluons are determined by strong interaction, which is characterized in the second term.

From Eq.~(\ref{cons intuitive eq}), we obtain the coefficients of $\frac{\delta M_{\rm u}}{M_{\rm u}}$ and $\frac{\delta M_{\rm d}}{M_{\rm d}}$ as
\begin{align}
\label{intuitive cons result}
\frac{m_{\rm u}}{M_{\rm u}}\frac{\partial M_{\rm u}}{\partial m_{\rm u}} = \frac{m_{\rm u}}{M_{\rm u}}, \quad \quad \frac{m_{\rm d}}{M_{\rm u}}\frac{\partial M_{\rm u}}{\partial m_{\rm d}} = \frac{m_{\rm s}}{M_{\rm u}}\frac{\partial M_{\rm u}}{\partial m_{\rm s}} = 0, \quad \quad \frac{\Lambda_{\scriptscriptstyle\rm QCD}}{M_{\rm u}}\frac{\partial M_{\rm u}}{\partial \Lambda_{\scriptscriptstyle\rm QCD}} = 1- \frac{m_{\rm u}}{M_{\rm u}}, \nonumber 
\\
\frac{m_{\rm d}}{M_{\rm d}}\frac{\partial M_{\rm d}}{\partial m_{\rm d}} = \frac{m_{\rm d}}{M_{\rm d}}, \quad \quad \frac{m_{\rm u}}{M_{\rm d}}\frac{\partial M_{\rm d}}{\partial m_{\rm u}} = \frac{m_{\rm s}}{M_{\rm d}}\frac{\partial M_{\rm d}}{\partial m_{\rm s}} = 0, \quad \quad \frac{\Lambda_{\scriptscriptstyle\rm QCD}}{M_{\rm d}}\frac{\partial M_{\rm d}}{\partial \Lambda_{\scriptscriptstyle\rm QCD}} = 1- \frac{m_{\rm d}}{M_{\rm d}}. 
\end{align}
In calculating the above coefficients, we will use $\frac{m_{\rm u}}{m_{\rm d}} = 0.553$, the central value of $m_{\rm d} = 9.3 \, \text{MeV}$ in the modified minimal subtraction ($\overline{\text{MS}}$) scheme at a renormalization scale of $1 \, \text{GeV}$~\cite{Leutwyler 96}, and we will choose $M_{\rm u} = M_{\rm d} = 335 \, \text{MeV}$.

\subsubsection{Expressions for $M_{\rm u,d}$  with an explicit quark sea -- linear form}
\label{sec: linear NQM}

A method explicitly taking into account the sea quark contribution can be traced back to the internal structure of the constituent quarks~\cite{Fritzsch:2009xe}. Then, for $m_{\rm p} (\Lambda_{\scriptscriptstyle\rm QCD}, m_{\rm u}, m_{\rm d}, m_{\rm s})$ a linear realization of this method is
\beqar
\label{linear matching eq}
M_{\rm u} &=& a_{\rm lin} \Lambda_{\scriptscriptstyle\rm QCD} + b_{\rm u,lin} m_{\rm u} + b_{\rm d,lin} m_{\rm d} +b_{\rm s,lin} m_{\rm s} \, , \nonumber \\
M_{\rm d} &=& a_{\rm lin} \Lambda_{\scriptscriptstyle\rm QCD} + b_{\rm d,lin} m_{\rm u} + b_{\rm u,lin} m_{\rm d} +b_{\rm s,lin} m_{\rm s} \, , \qquad (B)
\eeqar
where we have related the coefficients in $M_{\rm d}$ with those in $M_{\rm u}$ following~\cite{Fritzsch:2009xe}. The coefficients $a_{\rm lin}$ and $b_{\rm q,lin}$'s are pure numbers. Each of the four terms of $M_{\rm u,d}$ can be obtained by inserting Eq.~(\ref{linear matching eq}) into an expression for the NQM based proton mass, $m_{{\rm p},\scriptscriptstyle{\rm NQM}}$, which we will discuss shortly (see e.g., Eq.~(\ref{zero order proton eq}), (\ref{hyperfine eq}) or (\ref{Isgur-Karl eq})). Then, applying the Hellmann-Feynman theorem 
\be
\label{HF theorem eq}
{\partial m_{{\rm p},\scriptscriptstyle{\rm NQM}} \over \partial m_{\rm q}} = {\partial m_{{\rm p},\scriptscriptstyle{\rm NQM}} \over \partial M_{\rm u}}{\partial M_{\rm u} \over \partial m_{\rm q}} + {\partial m_{{\rm p},\scriptscriptstyle{\rm NQM}} \over \partial M_{\rm d}}{\partial M_{\rm d} \over \partial m_{\rm q}} = B_{\rm q}  \;\;\;\; (q=u,d,s)  .
\ee
An example of the application of the Hellmann-Feynman theorem within the NQM is given in~\cite{HK ref}. From Eqs.~(\ref{linear matching eq}) and (\ref{HF theorem eq}), the coefficients of $\delta v_i / v_i$ ($v_i = m_{{\rm u,d,s}}, \Lambda_{\scriptscriptstyle\rm QCD}$) of $\delta M_{\rm u} / M_{\rm u}$ and $\delta M_{\rm d} / M_{\rm d}$ can be obtained as
\beqar
\label{linear realization result form eq}
\frac{m_{\rm u}}{M_{\rm u}} \frac{\partial M_{\rm u}}{\partial m_{\rm u}} &=& \frac{k_{\rm u} \, m_{\rm p} f_{T_{\rm u}} - k_{\rm d} \, m_{\rm p} f_{T_{\rm d}} \left(\frac{m_{\rm u}}{m_{\rm d}}\right)}{M_{\rm u} \left(k_{\rm u}^2 - k_{\rm d}^2 \right)} \, , \nonumber \\
\frac{m_{\rm d}}{M_{\rm u}} \frac{\partial M_{\rm u}}{\partial m_{\rm d}} &=& \frac{k_{\rm u} \, m_{\rm p} f_{T_{\rm d}} - k_{\rm d} \, m_{\rm p} f_{T_{\rm u}} \left(\frac{m_{\rm d}}{m_{\rm u}}\right)}{M_{\rm u} \left(k_{\rm u}^2 - k_{\rm d}^2 \right)} \, ,  \\
\frac{m_{\rm s}}{M_{\rm u}} \frac{\partial M_{\rm u}}{\partial m_{\rm s}} &=& \frac{m_{\rm p} f_{T_{\rm s}}}{M_{\rm u} \left(k_{\rm u} + k_{\rm d}\right)} \, ,  \nonumber \\
\frac{\Lambda_{\scriptscriptstyle\rm QCD}}{M_{\rm u}} \frac{\partial M_{\rm u}}{\partial \Lambda_{\scriptscriptstyle\rm QCD}} &=& 1 - \sum_{q=u,d,s} \frac{m_{\rm q}}{M_{\rm u}} \frac{\partial M_{\rm u}}{\partial m_{\rm q}} \, , \nonumber
\eeqar
where $k_{\rm u,d} = \frac{\partial m_{{\rm p},\scriptscriptstyle{\rm NQM}}}{\partial M_{\rm u,d}}$. The $\frac{v_i}{M_{\rm d}}\frac{\partial M_{\rm d}}{\partial v_i} \; (v_i = m_{{\rm u,d,s}}, \Lambda_{\scriptscriptstyle\rm QCD})$ are obtained from the corresponding $\frac{v_i}{M_{\rm u}}\frac{\partial M_{\rm u}}{\partial v_i}$ by switching $M_{\rm u} \leftrightarrow M_{\rm d}$ and $k_{\rm u} \leftrightarrow k_{\rm d}$. 

To get $k_{\rm u,d}$, we consider the following NQM based proton mass formulae as examples. 
To zeroth order, the proton mass is the sum of the masses of its two constituent $u$ quarks and one constituent $d$ quark
\be
\label{zero order proton eq}
m_{\rm p} = 2 M_{\rm u} + M_{\rm d} \, ,
\ee
so that 
\be
\label{intuitive ks}
k_{\rm u} = 2, \quad k_{\rm d} = 1. 
\ee
We will use $M_{\rm u} = M_{\rm d} = {1 \over 3} \, m_{\rm p}$ in Eq.~(\ref{linear realization result form eq}) when Eq.~(\ref{zero order proton eq}) is taken as the NQM based proton mass formula.  

Without some interaction between the constituent quarks, hadrons with the same constituent quark compositions would have a same mass, a phenomenon which is not observed in nature. To break the mass degeneracy, a spin-spin hyperfine term is introduced~\cite{De Rujula:1975ge}, and the resulting proton mass is
\be
\label{hyperfine eq}
m_{\rm p} = 2M_{\rm u} + M_{\rm d} + A' \left({1 \over 4 M_{\rm u}^2} - {1 \over M_{\rm u} M_{\rm d}} \right)  ,
\ee
where $A'$ is a constant usually determined to allow an optimal fit to the baryon octet and decuplet masses~\cite{constituent_bmm}. This spin-spin hyperfine term is commonly attributed to one-gluon exchange~\cite{De Rujula:1975ge}, or, in the chiral quark model~\cite{Manohar and Georgi}, it is explained as the interaction between the constituent quarks mediated by pseudoscalar mesons~\cite{chiQM ref}. Although interpreted with relating to different degrees of freedom (gluon or pseudoscalar mesons)~\cite{Li and Cheng}, this term nevertheless has a strong interaction origin, and therefore we will write the parameter $A'$ as $a_{\rm hyp} \Lambda_{\scriptscriptstyle\rm QCD}^3$, with $a_{\rm hyp}$ a pure dimensionless number. From this formula, we get 
\be
\label{textbook ks}
k_{\rm u} = 2 + A' \left(\frac{1}{M_{\rm d} M_{\rm u}^2} - \frac{1}{2 M_{\rm u}^3} \right), \quad k_{\rm d} = 1 + \frac{A'}{M_{\rm u} M_{\rm d}^2}. 
\ee 
We will use $M_{\rm u} = M_{\rm d} = 363 \,\text{MeV}$ and $A'=(298.05 \, \text{MeV})^3$ in Eq.~(\ref{linear realization result form eq}) when Eq.~(\ref{hyperfine eq}) is taken as the NQM based proton mass formula. Note that we have tuned $A'$ a bit compared to the value given in~\cite{constituent_bmm} to allow an exact fit to the proton mass.

Eq.~(\ref{hyperfine eq}) can be further refined by adding to it the kinetic term of the constituent quarks and a constituent quark mass independent term $M_0$, which represents the contributions of the confinement potential and the short-range color-electric interaction~\cite{Isgur:1978xj, HK ref}
\be
\label{Isgur-Karl eq}
m_{\rm p} = 2M_{\rm u} + M_{\rm d} + A'' \left({1 \over 4 M_{\rm u}^2} - {1 \over M_{\rm u} M_{\rm d}} \right) + B' \left(\frac{1}{M_{\rm u}} + \frac{1}{2M_{\rm d}} \right) + M_0 \, .
\ee
From the physical meaning of these two new terms, it may be reasonable to write the constants $B'$ and $M_0$ as $a_{\rm kin} \Lambda_{\scriptscriptstyle\rm QCD}^2$ and $a_{\rm cce} \Lambda_{\scriptscriptstyle\rm QCD}$, respectively, since the internal dynamics of a baryon is dominated by the strong interaction and the confinement is a strong interaction phenomenon. The constant $A''$ needs to be re-fit after introducing the two new terms, and we write it as $A''=a'_{\rm hyp} \Lambda_{\scriptscriptstyle\rm QCD}^3$. The parameters $a_{\rm kin}$, $a_{\rm cce}$ and $a'_{\rm hyp}$ are pure dimensionless numbers. We find that $k_{\rm u,d}$ from this formula are
\be
\label{Isgur-Karl ks}
k_{\rm u} = 2 + A'' \left(\frac{1}{M_{\rm d} M_{\rm u}^2} - \frac{1}{2 M_{\rm u}^3} \right) - \frac{B'}{M_{\rm u}^2}, \quad k_{\rm d} = 1+\frac{A''}{M_{\rm u} M_{\rm d}^2} - \frac{B'}{2 M_{\rm d}^2}. 
\ee
We will use $M_{\rm u} = M_{\rm d} = 335 \, \text{MeV}$, $A'' = (4^{\frac{1}{3}} 176.4 \, \text{MeV})^3$, $B' = (175.2 \, \text{MeV})^2$ and $M_0 = -57.4 \, \text{MeV}$ in Eq.~(\ref{linear realization result form eq}) when Eq.~(\ref{Isgur-Karl eq}) is taken as the NQM based proton mass formula. Note that we have tuned $M_0$ a bit compared to the value given in~\cite{HK ref} to allow an exact fit to the proton mass.

\subsubsection{$M_{\rm u,d}$ expressions with an explicit quark sea -- NJL model}
\label{sec: non-linear NQM}
As can be seen from Eq.~(\ref{linear realization result form eq}), the explicit inclusion of the sea quark contribution in the linear form Eq.~(\ref{linear matching eq}) encodes the information of both the NQM based proton mass formula and the $f_{T_{\rm q}}$'s. However, different realizations from the NQM alone are also possible. Moreover, although the constituent quark masses are usually taken as fitting parameters in the study of hadron properties, it is certainly more illuminating if some concrete physical origin of these quantities can be given and encoded in their mass formulae. As suggested in~\cite{Manohar and Georgi, chiSSB ref}, the constituent quark masses are closely related to 
spontaneous chiral symmetry breaking. An example of the constituent quark mass formulae applying this idea is given by the three flavor Nambu-Jona-Lasinio (NJL) model~\cite{NJL ref, NJL three flavor ref}, where the constituent quark masses are obtained from a set of gap equations
\beqar
\label{gap eq}
M_{\rm u} &=& m_{\rm u} - 2 g_s \Braket{\bar{u} u} - 2 g_{\scriptscriptstyle D} \Braket{\bar{d} d} \Braket{\bar{s} s}  , \nonumber \\
M_{\rm d} &=& m_{\rm d} - 2 g_s \Braket{\bar{d} d} - 2 g_{\scriptscriptstyle D} \Braket{\bar{u} u} \Braket{\bar{s} s}  , \\
M_{\rm s} &=& m_{\rm s} - 2 g_s \Braket{\bar{s} s} - 2 g_{\scriptscriptstyle D} \Braket{\bar{u} u} \Braket{\bar{d} d}  , \qquad (C)\nonumber 
\eeqar
where $\Braket{\bar{u} u}$, $\Braket{\bar{d} d}$ and $\Braket{\bar{s} s}$ are the quark condensates which are the order parameters of the spontaneous chiral symmetry breaking, and they are calculated by one loop integral 
\be
\Braket{\bar{u} u} = - i N_c \mathrm{Tr} \int \frac{d^4 p}{(2 \pi)^4} \frac{1}{\slashed{p} - M_{\rm u} + i \epsilon} \, , \nonumber
\ee
and $\Braket{\bar{d} d}$ ($\Braket{\bar{s} s}$) is obtained by changing $M_{\rm u}$ to $M_{\rm d}$ ($M_{\rm s}$). The $N_c$ is the number of colors, and we take it to be the real-world value 3. This integration can be performed by introducing a three-momentum cutoff $\Lambda_3$, and the result is 
\be
\label{gap loop integral}
\Braket{\bar{u} u} = - \frac{3}{2 \pi^2} M_{\rm u} \left[ \Lambda_3 \sqrt{\Lambda_3^2 + M_{\rm u}^2} - M_{\rm u}^2 \ln{\left(\frac{\Lambda_3 + \sqrt{\Lambda_3^2 + M_{\rm u}^2}}{M_{\rm u}}\right)}\right]  .
\ee
The $g_s$ and $g_{\scriptscriptstyle D}$ in Eq.~(\ref{gap eq}) are the coupling constants of the effective four-point and six-point interactions of the quark fields in the NJL Lagrangian, and they are fixed, together with $m_{\rm s}$ and the cutoff $\Lambda_3$, by the meson properties as explained in~\cite{HK ref, NJL three flavor ref}. We simply quote the result given in~\cite{HK ref} \footnote{We have tuned the values of $g_s$ and $g_{\scriptscriptstyle D}$ relative to the values given in~\cite{HK ref} to allow $M_{\rm u} = M_{\rm d} = 335 \, \text{MeV}$ and $M_{\rm s} = 527 \, \text{MeV}$ as exact solutions of Eq.~(\ref{gap eq}).}
\be
\label{parameter of NJL eq}
m_{\rm s} = 135.7 \, \text{MeV}, \;\; g_s \Lambda_3^2 = 3.65, \;\; g_{\scriptscriptstyle D} \Lambda_3^5 = -9.47,\;\; \Lambda_3 = 631.4 \, \text{MeV},
\ee
which we will use for our calculation. The other parameters we need in order to solve Eq.~(\ref{gap eq}) are the $u$ and $d$ current quark masses, which we take $m_{\rm u} = m_{\rm d} = 5.5 \, \text{MeV}$ following~\cite{HK ref}. Note that the form of Eq.~(\ref{gap eq}) requires $m_{\rm u} = m_{\rm d}$ if we assume $M_{\rm u} = M_{\rm d}$. The cutoff $\Lambda_3$ characterizes the spontaneous chiral symmetry breaking scale, while the latter is related to $\Lambda_{\scriptscriptstyle\rm QCD}$, as we explained in the paragraph below Eq.~(\ref{fTg expression eq}). Therefore, we will write $\Lambda_3 = a_{c,\scriptscriptstyle{\rm NJL}} \Lambda_{\scriptscriptstyle\rm QCD}$, $g_s = a_{s,\scriptscriptstyle{\rm NJL}} \Lambda_{\scriptscriptstyle\rm QCD}^{-2}$ and $g_{\scriptscriptstyle D} = a_{\scriptscriptstyle{D, \rm NJL}} \Lambda_{\scriptscriptstyle\rm QCD}^{-5}$, where the coefficients are pure dimensionless numbers. With these inputs, the constituent quark masses are solved from Eq.~(\ref{gap eq}), with the values $M_{\rm u} = M_{\rm d} = 335 \, \text{MeV}$ and $M_{\rm s} = 527 \, \text{MeV}$, and we get 
\begin{align}
\label{gap eq result}
&\frac{m_{\rm u}}{M_{\rm u}}\frac{\partial M_{\rm u}}{\partial m_{\rm u}} = \frac{m_{\rm d}}{M_{\rm d}}\frac{\partial M_{\rm d}}{\partial m_{\rm d}} = 0.0351,& \quad &\frac{m_{\rm d}}{M_{\rm u}}\frac{\partial M_{\rm u}}{\partial m_{\rm d}} = \frac{m_{\rm u}}{M_{\rm d}}\frac{\partial M_{\rm d}}{\partial m_{\rm u}}  = 0.0074,& \nonumber
\\
&\frac{m_{\rm s}}{M_{\rm u}}\frac{\partial M_{\rm u}}{\partial m_{\rm s}} = \frac{m_{\rm s}}{M_{\rm d}}\frac{\partial M_{\rm d}}{\partial m_{\rm s}}  =  0.0628 ,& \quad &\frac{\Lambda_{\scriptscriptstyle\rm QCD}}{M_{\rm u}}\frac{\partial M_{\rm u}}{\partial \Lambda_{\scriptscriptstyle\rm QCD}} =  \frac{\Lambda_{\scriptscriptstyle\rm QCD}}{M_{\rm d}}\frac{\partial M_{\rm d}}{\partial \Lambda_{\scriptscriptstyle\rm QCD}} = 0.8947.&
\end{align}

\subsubsection{Results and discussion of the NQM approach}

We can now calculate the dependence of the nucleon magnetic moments on $m_{\rm u,d,s}$ and $\Lambda_{\scriptscriptstyle\rm QCD}$, by 
\be
\label{constituent theory-experiment relation eq}
\frac{\delta g_{\scriptscriptstyle\rm NQM}}{g_{\rm exp}} = \frac{g_{\scriptscriptstyle\rm NQM}}{g_{\rm exp}} \frac{\delta g_{\scriptscriptstyle\rm NQM}}{g_{\scriptscriptstyle\rm NQM}} \, ,
\ee
where $g_{\rm exp}$ is the measured value of the $g$-factor, which equals $5.586$ for proton, and $-3.826$ for neutron~\cite{PDG ref}. The first term in the square bracket of $\frac{\delta g_{\scriptscriptstyle\rm NQM}}{g_{\scriptscriptstyle\rm NQM}}$ (Eq.~(\ref{constituent_general vary function eq})) is given in section~\ref{sec: fTqg}, while the second term in that square bracket can be obtained from section~\ref{sec: intuitive NQM}, \ref{sec: linear NQM} and \ref{sec: non-linear NQM} for each of the three different constituent quark mass models we have considered. The calculated coefficients of $\frac{\delta v_i}{v_i} \; (v_i = m_{\rm u,d,s}, \, \Lambda_{\scriptscriptstyle\rm QCD})$ of $\frac{\delta g_{\scriptscriptstyle\rm NQM}}{g_{\rm exp}}$ for proton and neutron are listed in Table~\ref{tab:NQM approach results table proton},  where the constituent quark mass formula used for each row is labeled as A, B or C, representing Eqs.~(\ref{cons intuitive eq}), (\ref{linear matching eq}) or (\ref{gap eq}), respectively, while the $1$, $2$, or $3$ following the label B represents Eq.~(\ref{zero order proton eq}), (\ref{hyperfine eq}) or (\ref{Isgur-Karl eq}), respectively. In all of the cases listed in 
Table~\ref{tab:NQM approach results table proton}, we use the $f_{T_{\rm q}}$ given in  Eq.
(\ref{fTqg result}).  Note that for case C, there is a slight inconsistency due to our choice of $m_{\rm u} = m_{\rm d}$, though this has only a minor numerical effect on the resulting $\kappa$'s.

\begin{table}[htb!]
\caption{The coefficients $\kappa_i$ of $\frac{\delta v_i}{v_i} \; 
(v_i = m_{\rm u,d,s}, \, \Lambda_{\scriptscriptstyle\rm QCD})$ in
$\frac{\delta g_{\scriptscriptstyle\rm NQM}}{g_{\rm exp}}$ for the proton (left)
and the neutron (right); see Eq.~(\ref{defkappa}) for their definition.}
\label{tab:NQM approach results table proton}
\begin{center}
\begin{tabular}{|l c c c c|}
\hline
      & $\kappa_{\rm u}$ \,\, & $\kappa_{\rm d}$  \,\,  & $\kappa_{\rm s}$ \,\, & $\kappa_{\scriptscriptstyle\rm QCD}$ \\
\hline\hline
A  & $ 0.013 $      & $0.036 $     & $0.36 $   & $-0.41 $ \\
B1  & $-0.0039 $   & $0.0070 $   & $0 $        & $-0.0031 $ \\
B2  & $ 0.0029 $   & $0.021  $     & $0.11 $   & $-0.13 $ \\
B3  & $-0.0029 $   & $0.022  $    & $0.070 $ & $-0.089 $ \\
C  & $-0.0050$     & $0.029 $     & $0.30 $   & $-0.32 $ \\ 
\hline
\end{tabular}
\begin{tabular}{|l c c c c|}
\hline
      & $\kappa_{\rm u}$ \,\, & $\kappa_{\rm d}$  \,\,  & $\kappa_{\rm s}$ \,\, & $\kappa_{\scriptscriptstyle\rm QCD}$ \\
\hline\hline
A  & $0.021$    & $0.020$          & $0.35$        & $-0.40$ \\
B1  & $0.0056$  & $-0.010$       & $0$              & $0.0045$ \\
B2  & $0.012$    & $0.0033$       & $0.11$        & $-0.12$ \\
B3  & $0.011$    & $-0.0043$      & $0.068$      & $-0.075$ \\
C  & $0.010$    & $0.013$        & $0.29$        & $-0.32$ \\ 
\hline
\end{tabular}
\end{center}
\end{table}

The coefficients in Table~\ref{tab:NQM approach results table proton} show a relatively strong dependence on the constituent quark mass models used. Most of the coefficients in A and C are closer and much larger compared to their corresponding values in B. While the first and the second terms in Eq.~(\ref{constituent_general vary function eq}) are independent of each other for A and C, 
the same $f_{T_{\rm q}}$'s appear in both terms of Eq.~(\ref{constituent_general vary function eq}) for B, as can be seen from Eq.~(\ref{linear realization result form eq}) and thus these two terms are largely canceled due to a relative sign. We can also see a relatively strong dependence of the coefficients on the NQM based proton mass formulae when comparing the rows B1, B2 and B3. 

\begin{figure}
\begin{center}
\epsfig{file=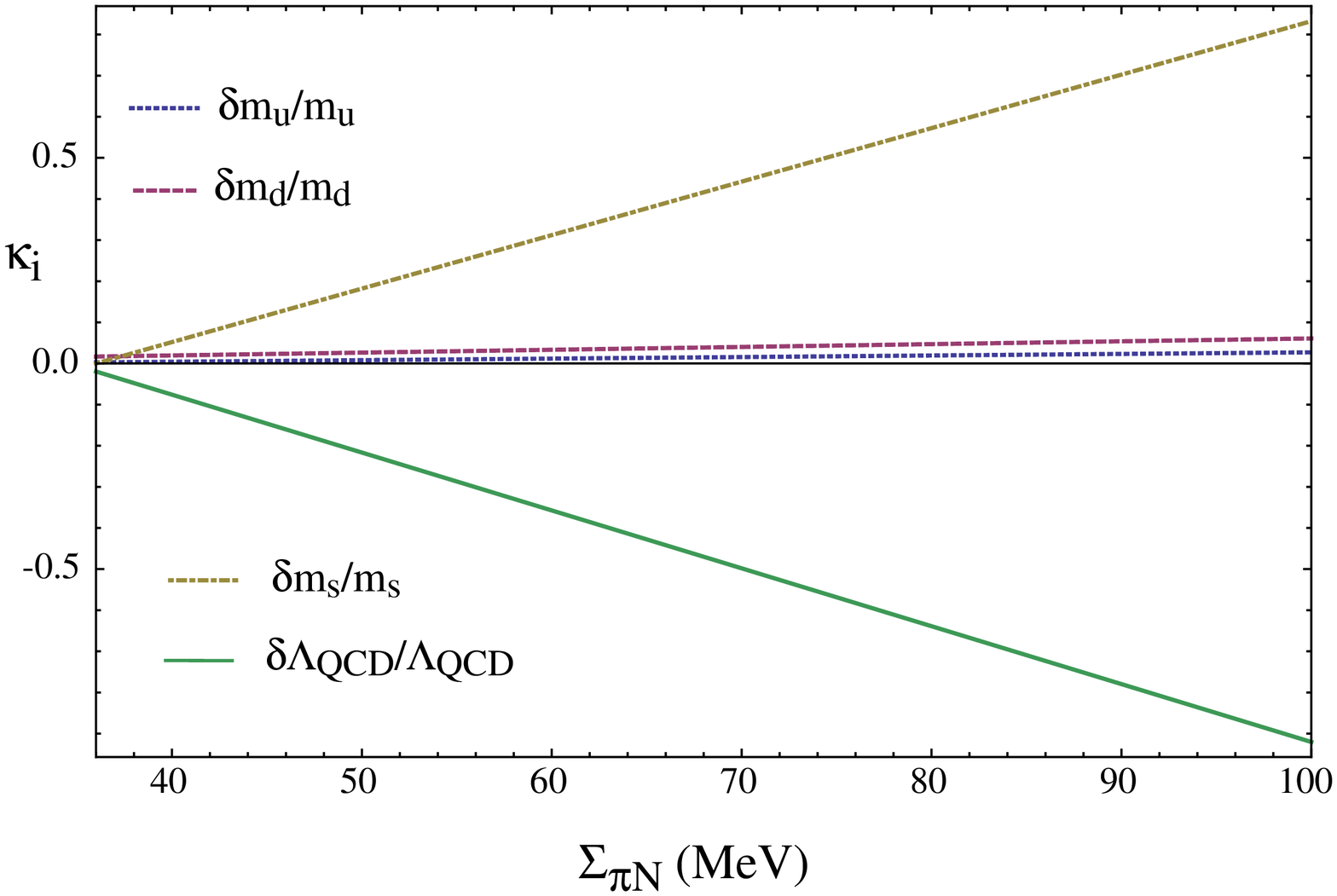,width=7.5cm}
\epsfig{file=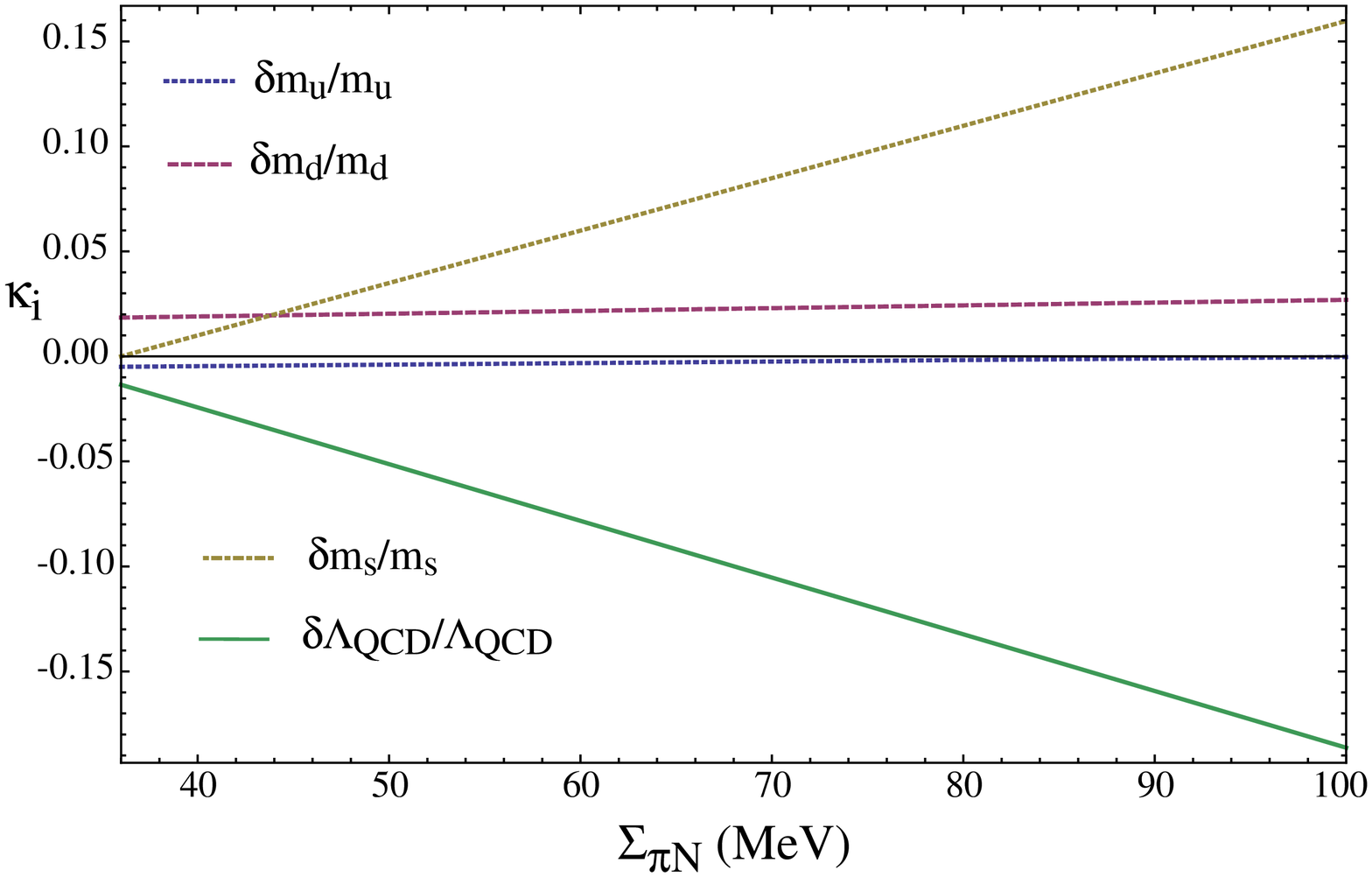,width=7.5cm}
\end{center}
\caption{The dependence of the coefficients in A for  the proton on $\SigmapiN$ (left)
and of the coefficients in B3 (right).}
\label{fig: AT vs SigmapiN}
\end{figure}

Furthermore, there is an uncertainty in the coefficients listed in Table~\ref{tab:NQM approach results table proton} due to the uncertainty of the $\pi$-nucleon sigma term $\SigmapiN$. A discussion of the impact of the uncertainty of $\SigmapiN$ on the interpretations of experimental searches for dark matter can be found in~\cite{sigma term ref2}. We plot the dependence of the coefficients in row A for the proton on $\SigmapiN$ in the left panel of Fig.~\ref{fig: AT vs SigmapiN}. A similar plot of the coefficients for the proton in row B3
is given in the right panel of Fig.~\ref{fig: AT vs SigmapiN}. As can be seen from these plots, the coefficients 
of $\delta m_{\rm s}/m_{\rm s}$ and $\delta \Lambda_{\scriptscriptstyle\rm QCD}/\Lambda_{\scriptscriptstyle\rm QCD}$ show a strong dependence on the value of $\SigmapiN$. Therefore it is important to pin down the value of $\SigmapiN$ if this quantity is used in the study of the current quark mass and $\Lambda_{\scriptscriptstyle\rm QCD}$ dependence of the proton $g$-factor. The same conclusion applies for the neutron $g$-factor, for which the behavior of the plots are similar to that shown in Fig.~\ref{fig: AT vs SigmapiN}. 

In addition to the relatively strong dependence of the $\kappa_i$ on the proton and constituent quark mass formula as well as the value of $\SigmapiN$ we have discussed above, some other comments for this NQM approach in the study of the dependence of the nucleon $g$-factors on the fundamental constants are in order. 
Our assumption that the various  parameters in the constituent quark mass formulae and the NQM based proton mass formulae take power law forms for $\Lambda_{\scriptscriptstyle\rm QCD}$ may be valid only in the chiral limit. Therefore, some current quark mass dependence may be lost and the $\Lambda_{\scriptscriptstyle\rm QCD}$ dependence may not be very accurately determined from these formulae. To get a more accurate dependence, one may also wish to consider relativistic corrections~\cite{relativisitic corrections ref} and/or corrections based on higher-dimension terms in the chiral quark model~\cite{Manohar and Georgi} for Eq.~(\ref{constituent_starting eq}), and then the dependence on the current quark masses and $\Lambda_{\scriptscriptstyle\rm QCD}$ will change correspondingly. Finally, in the above analysis, we did not consider the electromagnetic contribution to the proton mass or the constituent quark mass formulae, and thus we may have missed some dependence on the fine structure constant in this approach.

\subsection{The chiral perturbation theory approach}

The second approach we consider is $\chi$PT, which provides a systematic method of addressing the low energy properties of the hadrons~\cite{chipt classical, chipt pedagogical review}. In contrast to the strong model-dependent NQM approach we considered in the previous subsection, $\chi$PT can give model-independent calculations of the nucleon magnetic moments within a perturbative field theory framework in terms of the hadronic degrees of freedom. However, as we will see, our goal of extracting the current quark mass and $\Lambda_{\scriptscriptstyle\rm QCD}$ dependence is limited by our lack of knowledge of the accurate values of the coupling constants, the so called low energy constants (LECs), appearing in the effective Lagrangians of $\chi$PT. These Lagrangians, and the Feynman diagrams generated by them, are organized according to a power counting scheme, and the number of LECs we will have to deal with increases as we include higher order contributions to the nucleon magnetic moments. 

By construction, the LECs in the SU(3) $\chi$PT which we will consider do not depend on the light quark ($u$, $d$ and $s$) masses, and they should in principle be calculable in terms of the heavy quark ($c$, $b$ and $t$) masses and $\Lambda_{\scriptscriptstyle\rm QCD}$. Without the ability to solve non-perturbative QCD, the LECs are usually determined by fitting to experimental data for the pertinent physical observables, or estimated theoretically by QCD-inspired models and some other approaches (e.g., the resonance saturation method), and they can also be fixed by lattice calculations (for a discussion of the LECs, see for example,~\cite{chipt LECs} and the references therein). Most of the LECs are renormalization scale dependent in such a way that they cancel the renormalization scale dependent loop integrals so that the final results for the physical observables are renormalization scale independent. Furthermore, the values of the LECs are expected to be given by dimensional analysis~\cite{Manohar and Georgi,dim analysis} up to numerical factors of order one. Since the two quantities involved in such an analysis, namely, the Goldstone boson decay constant (for the meson octet) and the typical mass of the light but non-Goldstone states, are both pure numbers times $\Lambda_{\scriptscriptstyle\rm QCD}$ in the chiral limit, we will assume that all the LECs under discussion are functions of $\Lambda_{\scriptscriptstyle\rm QCD}$ and the renormalization scale, and by this assumption we neglect the heavy quark mass dependence in the LECs.  

For $\chi$PT in the meson-baryon sector, needed for the calculations of nucleon magnetic moments, there exist several renormalization schemes in the literature to ensure consistent power counting which is troubled by the introduction of the baryon mass as a new scale which is non-vanishing in the chiral limit. Among these renormalization schemes, the most studied in the early days in the calculations of octet baryon magnetic moments is the heavy baryon chiral perturbation theory (HB$\chi$PT) approach~\cite{HBchiPT ref}. Due to a strong cancellation between the leading order $\mathcal{O}(q^2)$ and the next-to-leading order $\mathcal{O}(q^3)$ results for this approach ($q$ denotes external momentum in the power counting scheme), one is forced to consider still higher order contributions. We will consider the results for this approach to order $\mathcal{O}(q^4)$~\cite{HBchiPT mm ref1, HBchiPT mm ref2, HBchiPT mm ref3, HBchiPT mm ref4}, with (HBwD) and without (HBw/oD) the explicit inclusion of the baryon decuplet states in loops. We will also consider a result from a more recently developed extended-on-mass-shell (EOMS) renormalization scheme~\cite{EOMS ref}, which gives more convergent results at $\mathcal{O}(q^3)$ without~\cite{EOMS mm ref1} or with~\cite{EOMS mm ref2} the inclusion of decuplet states in loops.  
We will restrict our attention to the EOMS without decuplets to avoid the introduction of 
several new parameters which do not improve the convergence.  

At leading order, the octet baryon magnetic moments can be calculated from the Feynman diagrams of chiral order $\mathcal{O}(q^2)$, and the results for both the HB$\chi$PT and EOMS approaches have the same expressions as linear combinations of two LECs $\mu_D$ and $\mu_F$, 
\be
\label{chiPT leading order}
\mu_B^{(2)} \equiv \alpha_B = \alpha_B^D \mu_D + \alpha_B^F \mu_F \, , 
\ee
where $\alpha_{\rm p}^D = 1/3$ and $\alpha_{\rm p}^F = 1$ for the proton, and $\alpha_{\rm n}^D = -2/3$ and $\alpha_{\rm n}^F = 0$ for the neutron. Note that we are writing down the magnetic moments (rather than the anomalous magnetic moments) directly in units of $\mu_{\mathrm{N}}$, and therefore the $\mu_F$ value we use may differ by $1$ compared to the value given in some of the references. 

At $\mathcal{O}(q^3)$ and higher order, the results of HB$\chi$PT and EOMS differ. For the HBw/oD approach, we use the result given in~\cite{HBchiPT mm ref1, HBchiPT mm ref2, HBchiPT mm ref3, HBchiPT mm ref4}, essentially using the notation of \cite{HBchiPT mm ref3}. At $\mathcal{O}(q^3)$ it is
\be
\label{HBchiPT order 3 w/o decuplet}
\mu_B^{(3)} = \sum_{X=\pi, K} \beta_B^X \frac{m_{\rm p} M_X}{8 \pi f_X^2},
\ee
where 
\be
\beta_{\rm p}^\pi = - (D+F)^2, \quad \beta_{\rm p}^K = -{2 \over 3} (D^2 +3 F^2),  \quad \beta_{\rm n}^\pi = (D+F)^2,  \quad \beta_{\rm n}^K = - (D - F)^2. \nonumber
\ee
$D$ and $F$ are dimensionless LECs. We will use the empirical values $m_{\rm p} = 938.3 \, \text{MeV}$, the pion decay constant $f_\pi = 93 \, \text{MeV}$ and the kaon decay constant $f_K = 1.2 f_\pi$ in our calculations, but we will consider these quantities take the forms of their corresponding LECs, namely, the average octet baryon mass, $m_0$ and the Goldstone boson decay constant $F_0$, since the differences between these quantities and their corresponding LECs give contributions to the octet baryon magnetic moments beyond $\mathcal{O}(q^4)$ which is the highest order we will consider. Therefore, in contrast to the NQM approach we considered in the previous subsection, we do not need to vary the proton mass in this subsection in calculating the current quark masses and $\Lambda_{\scriptscriptstyle\rm QCD}$ dependence of the nucleon magnetic moments. For the same reason, we take the pion mass, $M_\pi$, and kaon mass, $M_K$, at their empirical values of $138 \, \text{MeV}$ and $495 \, \text{MeV}$, respectively, in our calculations, while we only take their lowest order forms $M_\pi = (2 B_0 \hat{m})^{1/2}$ and $M_K = [B_0 (\hat{m} + m_{\rm s})]^{1/2}$, where the $B_0$ is an LEC with mass dimension, when considering light current quark masses and $\Lambda_{\scriptscriptstyle\rm QCD}$ dependences. We will work in the isospin-symmetric limit in this and the next subsection such that $m_{\rm u} = m_{\rm d} \equiv \hat{m}$. The expression for the $\eta$ mass, $M_\eta = [2/3 B_0 (\hat{m} + 2 m_{\rm s})]^{1/2}$ is also needed at $\mathcal{O}(q^4)$ in the HB$\chi$PT approach and at $\mathcal{O}(q^3)$ in the EOMS approach, and we use its empirical value of $548 \, \text{MeV}$.

For the case HBwD, we have the following terms in addition to Eq.~(\ref{HBchiPT order 3 w/o decuplet}) \cite{HBchiPT mm ref1, HBchiPT mm ref3, HBchiPT mm ref4},
\be
\mu_B^{(3)} = \sum_{X= \pi, K} \frac{m_{\rm p}}{8 \pi f_X^2} F (M_X, \delta, \lambda) \beta_B^{\prime X} \, ,
\label{HBwDq3addition}
\ee
where 
\be
\pi F(M, \delta, \lambda) = - \delta \, {\rm ln} \frac{M^2}{\lambda^2} 
+ 
	\left\{
    \begin{array}{ll}
        2\sqrt{M^2-\delta^2} \, [ \pi/2 - {\rm arctan} \,
          (\delta / \sqrt{M^2-\delta^2})], & M \geqslant  \delta ,\\ 
	\\
         - 2\sqrt{\delta^2-M^2} \, {\rm ln}
      [(\delta +\sqrt{\delta^2-M^2}) / M], & M < \delta,
    \end{array}
 \right. \nonumber
\ee
and
\be
\beta_{\rm p}^{\prime \pi} = -\frac{2}{9} \mathcal{C}^2,  \quad \beta_{\rm p}^{\prime K} = \frac{1}{18} \mathcal{C}^2,  \quad\beta_{\rm n}^{\prime \pi} = \frac{2}{9} \mathcal{C}^2,  \quad  \beta_{\rm n}^{\prime K} = \frac{1}{9} \mathcal{C}^2, \nonumber
\ee
where $\mathcal{C}$ is a dimensionless LEC, $\lambda$ is the renormalization scale in dimensional regularization, and $\delta$ is the decuplet-octet mass splitting for which we take to be a number times $\Lambda_{\scriptscriptstyle\rm QCD}$ with a value of $300 \, \text{MeV}$.  

At $\mathcal{O}(q^4)$ in the HB$\chi$PT approach, more LECs appear in the results and the formulae become lengthy. For the case HBw/oD, we take~\cite{HBchiPT mm ref2} \footnote{There is a misprint in the third term of $\mu_B^{(4,d+e+f)}$ in~\cite{HBchiPT mm ref2}, where the sign in front should be `$-$'.}
\be
\mu_B^{(4)} = \mu_B^{(4,c)} +\mu_B^{(4,d+e+f)}+\mu_B^{(4,g)}+\mu_B^{(4,h+i)}+\mu_B^{(4,j)},
\label{HBw/oDq4}
\ee
with
\be
\mu_{\rm p}^{(4,c)} = a_3 + a_4 + \frac{1}{3} a_5 + {1 \over 3} a_6 - {1 \over 3} a_7, \quad
\mu_{\rm n}^{(4,c)} = - \frac{2}{3} a_5 - {2 \over 3} a_6 - {1 \over 3} a_7, \nonumber
\ee
and
\beqar
\mu_B^{(4,d+e+f)} = && \sum_{X=\pi,K} \delta_B^X \frac{M_X^2}{16 \pi^2 f_X^2} {\rm ln}\frac{M_X}{\lambda} + 
  \sum_{X=\pi,K, \eta} \eta_B^X \frac{M_X^2}{16 \pi^2 f_X^2} \left({\rm ln}\frac{M_X}{\lambda} + 1 \right) \nonumber \\ 
\nonumber \\
  && - \left[  \sum_{X=\pi,K,\eta} \phi_B^X \frac{M_X^2}{8 \pi^2 f_X^2} \left(3 \, {\rm ln}\frac{M_X}{\lambda} + 1\right)\right] \alpha_B, \nonumber
\eeqar
with 
\beqar
&&\delta_{\rm p}^\pi = -\mu_D - \mu_F, \quad \delta_{\rm p}^K = -2 \mu_F, \quad \delta_{\rm n}^\pi = \mu_D + \mu_F, \quad \delta_{\rm n}^K = \mu_D - \mu_F, \nonumber \\
&&\eta_{\rm p}^\pi = \frac{1}{2} (D+F)^2 (\mu_D - \mu_F), \quad \eta_{\rm p}^K = -(\frac{1}{9} D^2 - 2 D F + F^2) \mu_D - (D-F)^2 \mu_F, \nonumber \\
&&\eta_{\rm p}^\eta = - \frac{1}{18} (D-3F)^2 (\mu_D + 3 \mu_F), \quad \eta_{\rm n}^\pi = - (D+F)^2 \mu_F, \nonumber \\
&&\eta_{\rm n}^K = (-\frac{7}{9} D^2 + \frac{2}{3} D F + F^2)\mu_D + (D-F)^2 \mu_F, \quad \eta_{\rm n}^\eta = \frac{1}{9} (D-3F)^2 \mu_D, \nonumber \\
&&\phi^\pi_{\rm p,n} = \frac{3}{4} (D+F)^2, \quad \phi^K_{\rm p,n} = \frac{5}{6}D^2 -D F +\frac{3}{2} F^2, 
\quad \phi_{\rm p,n}^\eta = \frac{1}{12} (D-3F)^2, \nonumber 
\eeqar
and 
\be
\mu_B^{(4,g)} = \sum_{X=\pi,K} \gamma^X_B \frac{m_{\rm p} M_X^2}{4 \pi^2 f^2_X} \, {\rm ln} \frac{M_X}{\lambda}, \nonumber
\ee
with
\beqar
&&\gamma^\pi_{\rm p} = 2 a_{9} + 2 \left(a_{10} + \frac{1}{8 m_{\rm p}}\right), \quad 
 \gamma^K_{\rm p} = a_{8} + 4 \left(a_{10} + \frac{1}{8 m_{\rm p}}\right), \nonumber \\
&&\gamma^\pi_{\rm n} = -2 a_{9} - 2 \left(a_{10} + \frac{1}{8 m_{\rm p}}\right), \quad 
\gamma^K_{\rm n} = -2 a_{9} + 2 \left(a_{10} + \frac{1}{8 m_{\rm p}}\right), \nonumber 
\eeqar
and 
\be
\mu_B^{(4,h+i)} = \sum_{X=\pi,K} \beta_B^X \frac{M_X^2}{16 \pi^2 f_X^2} \left(5 \, {\rm ln} \frac{M_X}{\lambda} +1 \right), \nonumber
\ee
and
\be
\mu_B^{(4,j)} = - \sum_{X=\pi, K} \theta_B^X \frac{m_{\rm p}}{2 \pi^2 f^2_X} \left(2\, {\rm ln}\frac{M_X}{\lambda} + 1\right), \nonumber 
\ee
with
\beqar
&&\theta_{\rm p}^\pi = (D+F)^2 \left[M_K^2 a_{11} + \left(M_\pi^2 - M_K^2\right) a_{12} \right], \quad
 \theta_{\rm p}^K = \frac{1}{6} \left[ \left(3 F + D\right)^2 M_\eta^2 + 3 (D-F)^2 M_\pi^2\right] a_{11}, \nonumber \\
&&\theta_{\rm n}^\pi = -(D+F)^2 \left[M_K^2 a_{11} + (M_\pi^2 - M_K^2) a_{12} \right], \quad
 \theta_{\rm n}^K = (D-F)^2 M_\pi^2 a_{11}. \nonumber
\eeqar
In the above formulae, the LECs $a_{8,9,10,11,12}$ (labeled $b_{9,10,11,D,F}$ in~\cite{HBchiPT mm ref2}) with their values in units of $\text{GeV}^{-1}$ are $a_8 = 0.81$, $a_{9} = 0.95$, $a_{10} = 0.36$, $a_{11} = - 0.192$ and $a_{12} = - 0.210$, where the first three are estimated by the resonance saturation method which takes into account the contribution from the baryon decuplet while the other two are determined by fitting to the baryon octet masses, as explained in detail in \cite{HBchiPT mm ref2}. We take the value of the $\eta$ decay constant to be $f_\eta = 1.2 f_\pi$, but we will consider it taking the form of its corresponding LEC, $F_0$, for the same reason explained above for the other Goldstone boson decay constants.

At this order, an uncertainty arises for our extraction of the dependence of magnetic moments on the current quark masses and $\Lambda_{\scriptscriptstyle\rm QCD}$ due to the uncertainties of the values of the LECs appearing in the counter term Lagrangian, and they are denoted as $a_{3,4,5,6,7}$ in~Eq. (\ref{HBw/oDq4}). These dimensionless numbers  have actually absorbed the light current quark masses in the counter term Lagrangian in contrast to their corresponding true LECs which are independent of the light current quark masses. Therefore these redefined LECs should contain a factor $m_{\rm s} / \Lambda_{\scriptscriptstyle\rm QCD}$, if we neglect the contributions from $\hat{m}$ as its value is much smaller than $m_{\rm s}$. Two other LECs in the counter term Lagrangian are also present at this order, and they are combined with the two LECs appearing in the $\mathcal{O}(q^2)$ result Eq.~(\ref{chiPT leading order}), as 
\be
\label{redefine dimension 2}
\mu_{D,F} \rightarrow \mu_{D,F} + 4 B_0 (2 \hat{m} + m_{\rm s}) \tilde{\mu}_{D,F} \, , 
\ee
where $\tilde{\mu}_{D,F}$ are LECs appearing in $\mathcal{O}(q^4)$ counter term Lagrangian. Then all seven of these redefined LECs, $\mu_{D,F}$ and $a_{3,4,5,6,7}$, are used as fitting parameters to perform an exact fit to the seven available octet baryon magnetic moments. Since they are used as fitting parameters, and indeed different values for them are obtained with and without the explicit inclusion of baryon decuplet states in loops, as well as when  different values of other LECs are used for the fittings (see the discussion below), it is hard to get an accurate extraction of the light current quark mass and $\Lambda_{\scriptscriptstyle\rm QCD}$ dependence from these redefined LECs. For the light current quark masses dependence, we will only consider the $m_{\rm s}$ dependence for $a_{3,4,5,6,7}$, while we will not try to extract such dependence for the redefined $\mu_{D,F}$ (denoted as $a_{1,2}$ in~\cite{HBchiPT mm ref3}), since we do not know the relative size of the two terms on the right hand side of Eq.~(\ref{redefine dimension 2}), where only one of the two terms has the light current quark mass dependence, although such dependence in these two redefined LECs may be not small, as suggested in~\cite{HBchiPT mm ref2} when comparing the fitting values up to $\mathcal{O}(q^3)$ with the ones up to $\mathcal{O}(q^4)$. We take the values $\mu_D = 3.71$, $\mu_F = 3.25$, $a_3 = - 0.50$, $a_4 = - 0.15$, $a_5 = - 0.25$, $a_6 = 0.50$ and $a_7= - 0.46$ given in~\cite{HBchiPT mm ref2}, where $F=0.5$, $D=0.75$ is used, and the renormalization scale $\lambda$ is taken to be $0.8 \, \text{GeV}$. 

For HBwD at $\mathcal{O}(q^4)$, we take~\cite{HBchiPT mm ref3} \footnote{There is a misprint in the form of the $ \pi G(M,\delta, \lambda)$ for the case $M\geqslant \delta$, where the sign in front should be `$-$'.}
\beqar
\label{HBwDq4}
\mu_B^{(4)} = &&\mu_B^{(4,c)}
+ \sum_{X=\pi, K, \eta} \frac{1}{32 \pi^2 f_X^2} \, (\gamma_B^{\prime X} - 2 \, \phi_B^{\prime X} \alpha_B)\,  M_X^2 \, {\rm ln} \frac{M^2_X}{\lambda^2} \nonumber \\
\nonumber \\
&& + \sum_{X=\pi, K, \eta} \frac{1}{32 \pi^2 f_X^2} \left[(\tilde{\gamma}_B^{\prime X} - 2 \, \tilde{\phi}_B^{\prime X} \alpha_B) L_{(3/2)} (M_X, \delta, \lambda) + \hat{\gamma}_B^{\prime X} L_{(3/2)}^\prime (M_X, \delta, \lambda) \right], 
\eeqar
where
\be
L_{(3/2)} (M, \delta, \lambda) = M^2 \, {\rm ln} \frac{M^2}{\lambda^2} + 2  \pi  \delta \,  F(M, \delta, \lambda), \nonumber
\ee
and 
\be
L_{(3/2)}^\prime (M, \delta, \lambda)  = M^2 \, {\rm ln} \frac{M^2}{\lambda^2} + \frac{2 \pi}{3 \delta} G(M,\delta,\lambda), \nonumber
\ee
with
\be
\pi G(M, \delta, \lambda) = - \delta^3 \, {\rm ln} \frac{M^2}{\lambda^2} + \pi M^3
+ 
	\left\{
    \begin{array}{ll}
        - 2 (M^2-\delta^2)^{3/2} \, [ \pi/2 - {\rm arctan} \,
          (\delta / \sqrt{M^2-\delta^2})], & M \geqslant  \delta ,\\ 
	\\
         - 2 (\delta^2-M^2)^{3/2} \, {\rm ln}
      [(\delta +\sqrt{\delta^2-M^2}) / M], & M < \delta,
    \end{array}
 \right. \nonumber
\ee
and
\beqar
&&\tilde{\gamma}_{\rm p}^{\prime \pi} = \frac{80}{27} \mathcal{C}^2 \mu_C, \quad \tilde{\gamma}_{\rm p}^{\prime K} = \frac{10}{27} \mathcal{C}^2 \mu_C, \quad \tilde{\gamma}_{\rm p}^{\prime \eta} = 0, \nonumber \\
&&\tilde{\gamma}_{\rm n}^{\prime \pi} = -\frac{20}{27} \mathcal{C}^2 \mu_C, \quad 
\tilde{\gamma}_{\rm n}^{\prime K} = - \frac{10}{27} \mathcal{C}^2 \mu_C, \quad 
\tilde{\gamma}_{\rm n}^{\prime \eta} = 0, \nonumber  \\
&&\hat{\gamma}_{\rm p}^{\prime \pi} = \frac{8}{9} \mathcal{C} (D+F) \mu_T, \quad 
\hat{\gamma}_{\rm p}^{\prime K} = \frac{2}{9} \mathcal{C} (3 D- F) \mu_T, \quad 
\hat{\gamma}_{\rm p}^{\prime \eta} = 0, \nonumber \\
&&\hat{\gamma}_{\rm n}^{\prime \pi} = - \frac{8}{9} \mathcal{C} (D+F) \mu_T, \quad 
\hat{\gamma}_{\rm n}^{\prime K} = - \frac{4}{9} \mathcal{C} F \mu_T, \quad 
\hat{\gamma}_{\rm n}^{\prime \eta} = 0, \nonumber \\
&&\tilde{\phi}_{\rm p,n}^{\prime \pi} = 2 \mathcal{C}^2, \quad 
\tilde{\phi}_{\rm p,n}^{\prime K} = \frac{1}{2} \mathcal{C}^2, \quad
\tilde{\phi}_{\rm p,n}^{\prime \eta} = 0. \nonumber 
\eeqar
The other coefficients are related to the ones given in Eq.~(\ref{HBw/oDq4}), as
\be
\gamma_B^{\prime \pi,K} = \delta_B^{\pi,K} + \eta_B^{\pi,K}, \quad \gamma_B^{\prime \eta} = \eta_B^{\eta}, \quad  \phi_B^{\prime X} = 3 \phi_B^X,  \nonumber
\ee
and $\mu_B^{(4,c)}$ is the same as the case of HBw/oD. 

In this case, since different LECs are used as inputs for the fittings in comparison to the HBw/oD case, the resulting fit values of the seven redefined LECs are different, and we take their values from Case (b) in Table~\textrm{II} of~\cite{HBchiPT mm ref3}, $a_1 = 3.946$, $a_2 = 2.353$, $a_3 = - 0.001$, $a_4 = - 0.172$, $a_5 = 0.569$, $a_6 = 0.694$ and $a_7 = - 1.165$, corresponding to the LECs inputs $F = 0.5$, $D = 0.75$, $\mathcal{C} = -1.5$, $\mu_T = -7.7$ and $\mu_C = 1.94$. A renormalization scale $\lambda = 1 \, \text{GeV}$ is used. 

For the EOMS approach, to minimize the number of LECs involved and thus perhaps the uncertainties introduced by them, we only consider the result given in Eq.~(2) to Eq.~(5) of~\cite{EOMS mm ref1} which does not include the baryon decuplet states in loops. The result is up to $\mathcal{O}(q^3)$, and the loop integrals are finite. The values $\mu_D = 3.82$ and $\mu_F = 2.20$
denoted as $\tilde{b}_6^D$ and $\tilde{b}_6^F$ in \cite{EOMS mm ref1}, after performing the EOMS scheme, are determined by a fit to minimize the $\tilde{\chi}^2 = \sum (\mu_{\rm th} - \mu_{\rm exp})^2$ as explained in that reference. 
For other quantities in the formula, $F_\phi = 1.17 f_\pi$ is the average of the physical values of $f_\pi$, $f_K$ and $f_\eta$, and we still use $M_B = 938.3 \, \text{MeV}$, $f_\pi = 93 \, \text{MeV}$, and the same values for $M_{\pi, K, \eta}$ as specified above. We take $D = 0.80$ and $F = 0.46$ as used in~\cite{EOMS mm ref1} for this EOMS approach.

We list the results of the two HB$\chi$PT and the one EOMS approaches in Table~\ref{tab:chiPT approaches for proton}, where we also need to specify the ratio of $m_{\rm s}$ to $\hat{m}$, for which we use $25$. Note that as we discussed above, we have assumed that the LECs ($a_{1,2,3,4,5,6,7}$, correspond to the  original LECs before the re-definition) depend only on $\Lambda_{\scriptscriptstyle\rm QCD}$ and the renormalization scale, $\lambda$, i.e.,
they do not depend on the light quark masses and there is no dependence on the 
renormalization scale in the full result. Therefore, the coefficient of $\delta \Lambda_{\scriptscriptstyle\rm QCD}/\Lambda_{\scriptscriptstyle\rm QCD}$ must be equal and opposite to the sum of the light quark mass contributions.

\begin{table}[ht!]
\caption{The coefficients, $\kappa_i$, of $\frac{\delta v_i}{v_i} \; (v_i = \hat{m}, \, m_{\rm s}, 
\, \Lambda_{\scriptscriptstyle\rm QCD})$ of $\frac{\delta g_{\scriptscriptstyle {\chi\rm PT}}}{g_{\rm exp}}$, 
defined as in Eq.~(\ref{defkappa}), for the proton (left) and the neutron (right).} 
\label{tab:chiPT approaches for proton}
\begin{center}
\begin{tabular}{|l c c c  |}
\hline
    & $2 \kappa_{\rm u} = 2 \kappa_{\rm d}$ \,\, & $\kappa_{\rm s}$ \,\, & $\kappa_{\scriptscriptstyle\rm QCD}$ \\
\hline\hline
HBw/oD  & $-0.050$    &  $-0.50$     & $0.54 $  \\
HBwD  & $0.034$ & $0.17 $          & $-0.21  $ \\
EOMS  & $-0.049 $   & $-0.031 $          & $0.080 $ \\
\hline
\end{tabular}
\begin{tabular}{| c c c|}
\hline
       $2 \kappa_{\rm u} = 2  \kappa_{\rm d}$ \,\, & $\kappa_{\rm s}$ \,\, & $\kappa_{\scriptscriptstyle\rm QCD}$ \\
\hline\hline
 $-0.16$    &  $-0.14$     & $0.30$  \\
 $-0.050$ & $0.32 $          & $-0.27  $ \\
 $-0.11 $   & $0.014 $          & $0.097 $ \\
\hline
\end{tabular}
\end{center}
\end{table}

We see from Table~\ref{tab:chiPT approaches for proton}, the numbers in each column differ considerably for nucleon magnetic moment formulae from different renormalization schemes and depend on the explicit inclusion of baryon decuplet states in loops. As we mentioned in the beginning of this subsection, we believe this discrepancy comes in a large part from our lack of knowledge of the accurate values of the LECs. In particular, many of the LECs involved in our calculations are used as fit parameters for the octet magnetic moments and masses, while their true values may be quite different from the numbers obtained by these fits. For example, if we use the values for Case (a) $F = 0.4$, $D=0.61$ and $\mathcal{C} = -1.2$ in~\cite{HBchiPT mm ref3}, the resulting values of the $a$'s also give an exact fit to the seven available octet baryon magnetic moments, with a prediction for the $\Sigma \Lambda$ transition moment similar to the one given by Case (b). However, one can see that many of the corresponding $a$'s for Case (a) and Case (b) differ greatly, and indeed, for Case (a) we get the coefficients from left to right of Table~\ref{tab:chiPT approaches for proton} as $0.014$, $0.14$ and $- 0.16$ for the proton, and $-0.039$, $0.26$ and $-0.23$ for the neutron, which are different from the results of Case (b). Therefore, it is crucial to pin down the values of LECs 
before one can make a better extraction of the light quark masses and $\Lambda_{\scriptscriptstyle\rm QCD}$ dependence in the $\chi$PT approach.

One can also estimate the dependence of $m_{\rm p}$ on the current quark masses and $\Lambda_{\scriptscriptstyle\rm QCD}$ from a formula for $m_{\rm p}$ within $\chi$PT. Such dependences can be used when one varies the electron-to-proton mass ratio, $\mu \equiv m_{\rm e} / m_{\rm p}$. However, as we explained at the end of the previous subsection, we prefer to use a common set of values for the coefficients of $\kappa_{\rm q}$ and $\kappa_{\rm QCD} $ of $\delta m_{\rm p} / m_{\rm p}$. Those values for the isospin-symmetric limit case are listed at the end of section~\ref{sec: fTqg}.

\subsection{The approach combining $\chi$PT and lattice QCD}

As another approach to study hadronic physics, lattice QCD provides a promising way to extract the current quark masses dependence of the nucleon magnetic moments, because one can do explicit calculations by assuming a sequence of different current quark masses in lattice computations, although in practice the computational cost is a limitation. Since most of the current lattice computations are still using input current quark masses much larger than their empirical values, an extrapolation of the lattice results to the physical point is needed. In the extrapolations for the physical observables, terms having non-analytic behaviors, $m_{\rm q}^{1/2}$ and $m_{\rm q} \log m_{\rm q}$, etc., which are predicted by $\chi$PT and have important contributions near the chiral limit, must be considered. 

An earlier study of this combined lattice and $\chi$PT approach for the nucleon magnetic moments uses an encapsulating form which is the Pad$\acute{\text{e}}$ approximant~\cite{Pade lattice},
\be
\mu_{\rm p,n} (M_\pi) = \frac{\mu_0}{1 - \frac{\chi_{\rm p,n}}{\mu_0} M_\pi + c M_\pi^2} \, ,
\ee
where $\chi_{\rm p,n}$ are fixed by the leading non-analytic term given by $\chi$PT, while $\mu_0$ and $c$ are allowed to vary to best fit the lattice data. 

A later development takes the finite range regulator (FRR)~\cite{FFR ref} as the regularization method rather than the traditional dimensional regularization for the results we discussed in the previous subsection, and the cut-off parameter in the FRR is a mass scale which can be interpreted as the inverse of the size of the nucleon. 

The current quark masses dependence for the nucleon magnetic moments is given in~\cite{lattice result ref 0402098}, and we simply quote the result there without going into any detail
\be
\label{lattice result eq1}
\frac{\delta g_{\rm p}}{g_{\rm p}} = - 0.087 \frac{\delta \hat{m}}{\hat{m}} - 0.013 \frac{\delta m_{\rm s}}{m_{\rm s}}, \,\,\, \frac{\delta g_{\rm n}}{g_{\rm n}} = - 0.118 \frac{\delta \hat{m}}{\hat{m}} + 0.0013 \frac{\delta m_{\rm s}}{m_{\rm s}}.
\ee
As the same argument we made for the $\chi$PT approach in the previous subsection, all parameters without light quark masses dependence are either pure numbers or are pure numbers time $\Lambda_{\scriptscriptstyle\rm QCD}$. Therefore, we obtain 
\beqar
\label{lattice result eq2}
\frac{\delta g_{\rm p}}{g_{\rm p}} &=& - 0.087 \frac{\delta \hat{m}}{\hat{m}} - 0.013 \frac{\delta m_{\rm s}}{m_{\rm s}} + 0.100 \frac{\delta \Lambda_{\scriptscriptstyle\rm QCD}}{\Lambda_{\scriptscriptstyle\rm QCD}}, \nonumber \\
\frac{\delta g_{\rm n}}{g_{\rm n}} &=& - 0.118 \frac{\delta \hat{m}}{\hat{m}} + 0.0013 \frac{\delta m_{\rm s}}{m_{\rm s}} + 0.1167 \frac{\delta \Lambda_{\scriptscriptstyle\rm QCD}}{\Lambda_{\scriptscriptstyle\rm QCD}}.
\eeqar

\section{Atomic clock constraints}

\subsection{Methodology}

As we have seen in section~\ref{Section2D}, the frequency shift is related to $\lbrace g_{\rm p}, g_{\rm n}, b, \mu,\alpha\rbrace$
by the relation
\begin{equation}\label{decomposition2}
 \frac{\dot\nu_{AB}}{\nu_{AB}} = \lambda_{g_{\rm p}}  \frac{\dot g_{\rm p}}{g_{\rm p}}
  +\lambda_{g_{\rm n}}  \frac{\dot g_{\rm n}}{g_{\rm n}}
 +\lambda_b  \frac{\dot b}{b}
  + \lambda_{\mu}  \frac{\dot\mu}{\mu} +\lambda_{\alpha}  \frac{\dot \alpha}{\alpha} ,
\end{equation}
where the coefficients $\lambda$ are given explicitly in Table~\ref{tab:coef1}. Then, in section~\ref{Section3},
we have expressed the dependence of the $g$-factors as
\begin{eqnarray}
\label{decomposition3 proton}
 \frac{\delta g_{\rm p}}{g_{\rm p}} &=& \kappa_{\rm u_p} \frac{\delta m_{\rm u}}{m_{\rm u}}
  +  \kappa_{\rm d_p} \frac{\delta m_{\rm d}}{m_{\rm d}} +
   \kappa_{\rm s_p} \frac{\delta m_{\rm s}}{m_{\rm s}} +
    \kappa_{\scriptscriptstyle\rm QCD_p} \frac{\delta \Lambda_{\scriptscriptstyle\rm QCD}}{\Lambda_{\scriptscriptstyle\rm QCD}}, \\
\label{decomposition3 neutron}
 \frac{\delta g_{\rm n}}{g_{\rm n}} &=& \kappa_{\rm u_n} \frac{\delta m_{\rm u}}{m_{\rm u}}
  +  \kappa_{\rm d_n} \frac{\delta m_{\rm d}}{m_{\rm d}} +
   \kappa_{\rm s_n} \frac{\delta m_{\rm s}}{m_{\rm s}} +
    \kappa_{\scriptscriptstyle\rm QCD_n} \frac{\delta \Lambda_{\scriptscriptstyle\rm QCD}}{\Lambda_{\scriptscriptstyle\rm QCD}},
\end{eqnarray}
where the coefficients $\kappa_i$ have been calculated for different models and collected in Tables~\ref{tab:NQM approach results table proton} and \ref{tab:chiPT approaches for proton} and Eq.~(\ref{lattice result eq2}), as well
as the dependence of the proton mass
\begin{equation}\label{decomposition4}
 \frac{\delta m_{\rm p}}{m_{\rm p}} = f_{T_{\rm u}} \frac{\delta m_{\rm u}}{m_{\rm u}}
  +  f_{T_{\rm d}} \frac{\delta m_{\rm d}}{m_{\rm d}} +
   f_{T_{\rm s}} \frac{\delta m_{\rm s}}{m_{\rm s}} +
   f_{T_{\rm g}} \frac{\delta \Lambda_{\scriptscriptstyle\rm QCD}}{\Lambda_{\scriptscriptstyle\rm QCD}} ,
\end{equation}
where the $f_{T_i}$ are given in Eqs.~(\ref{fTqg result}) and (\ref{fTqg isospin eq 2}). Also, following~\cite{flam, BFK2011}, $b$ depends on the quark mass and $\Lambda_{\scriptscriptstyle\rm QCD}$, and there it is found
\be
\label{b varying eq}
\frac{\delta b}{b} = \gamma_{\rm q} \frac{\delta \hat{m}}{\hat{m}} + \gamma_{\scriptscriptstyle\rm QCD} \frac{\delta \Lambda_{\scriptscriptstyle\rm QCD}}{\Lambda_{\scriptscriptstyle\rm QCD}}, 
\ee
with 
\be
\label{coefficients of b varying}
\gamma_{\rm q} = - \gamma_{\scriptscriptstyle\rm QCD} = -0.11. 
\ee

Assuming for simplicity that all Yukawa couplings are varying similarly,
i.e., $\delta h_i/h_i = \delta h/h$, the expansions~(\ref{decomposition3 proton}), (\ref{decomposition3 neutron}), (\ref{decomposition4}) and~(\ref{b varying eq}) can be inserted in Eq.~(\ref{decomposition2}) to
obtain
\begin{equation}\label{nufinal1}
 \frac{\dot\nu_{AB}}{\nu_{AB}} = \hat\lambda_{h}  \frac{\dot h}{h}
 + \hat\lambda_{v}\frac{\dot v}{v}
 + \hat\lambda_{\scriptscriptstyle\rm QCD} \frac{\dot \Lambda_{\scriptscriptstyle\rm QCD}}{\Lambda_{\scriptscriptstyle\rm QCD}}
 + \hat\lambda_{\alpha} \frac{\dot \alpha}{\alpha}.
\end{equation}
The coefficients $\hat\lambda$ are easily computed to be given by
\begin{eqnarray}
 \hat\lambda_{h} &=&  \lambda_{g_{\rm p}}(\kappa_{\rm u_p} + \kappa_{\rm d_p} + \kappa_{\rm s_p})
 + \lambda_{g_{\rm n}}(\kappa_{\rm u_n} + \kappa_{\rm d_n} + \kappa_{\rm s_n})
 + \lambda_b \gamma_{\rm q}
 +\lambda_\mu(1 -  f_{T_{\rm u}}  -  f_{T_{\rm d}} - f_{T_{\rm s}})\\
 \hat\lambda_{v}  &=& \hat\lambda_{h} \\
 \hat\lambda_{\scriptscriptstyle\rm QCD}  &=&  \lambda_{g_{\rm p}}\kappa_{\scriptscriptstyle\rm QCD_{\rm p}}
 + \lambda_{g_{\rm n}}\kappa_{\scriptscriptstyle\rm QCD_{\rm n}}
 + \lambda_b \gamma_{\scriptscriptstyle\rm QCD}
  - \lambda_\mu f_{T_{\rm g}}\\
 \hat\lambda_{\alpha}  &=& \lambda_{\alpha}.
\end{eqnarray}
The form~(\ref{nufinal1}) makes no assumption on unification and only relies
on the fact that all Yukawa couplings are varying in a similar way. It is important to note
here that the dimensionality constraint on the $f_{T_i}$, $\kappa_i$ and $\gamma_i$ parameters
implies that $\hat\lambda_{\scriptscriptstyle\rm QCD}=-\hat\lambda_{v}$ so that Eq.~(\ref{nufinal1}) actually
depends only on the combination of $X\equiv hv/\Lambda_{\scriptscriptstyle\rm QCD}$ and $\alpha$
as
\begin{equation}\label{nufinal1red}
 \frac{\dot\nu_{AB}}{\nu_{AB}} = \hat\lambda_{h}  \frac{\dot X}{X}
 + \hat\lambda_{\alpha} \frac{\dot \alpha}{\alpha}.
\end{equation}
This would not be the case if we had not assumed that $\delta h_i/h_i = \delta h/h$
for all Yukawa couplings.

Our first hypothesis concerning unification allows one to express the variation of the QCD scale
by means of Eq.~(\ref{DeltaLambda}) so that
\begin{eqnarray}\label{nufinal2}
 \frac{\dot\nu_{AB}}{\nu_{AB}} &=& \left(\hat\lambda_{h}+\frac29  \hat\lambda_{\scriptscriptstyle\rm QCD} \right) \frac{\dot h}{h}
 + \left(\hat\lambda_{v}+\frac29  \hat\lambda_{\scriptscriptstyle\rm QCD} \right) \frac{\dot v}{v}
 +\left( \hat\lambda_{\alpha} + R  \hat\lambda_{\scriptscriptstyle\rm QCD} \right)\frac{\dot \alpha}{\alpha} \nonumber \\
 &\equiv& H_S\left( \frac{\dot h}{h} + \frac{\dot v}{v} \right) + H_\alpha \frac{\dot \alpha}{\alpha}.
\end{eqnarray}
The second hypothesis on unification assumes that the variation of $v$ and $h$ are related by
Eq.~(\ref{enhance2}) so that
\begin{eqnarray}\label{nufinal3}
 \frac{\dot\nu_{AB}}{\nu_{AB}} &=& \left(\hat\lambda_{h}+\frac29  \hat\lambda_{\scriptscriptstyle\rm QCD} \right)(1+S) \frac{\dot h}{h}
 +\left( \hat\lambda_{\alpha} + R  \hat\lambda_{\scriptscriptstyle\rm QCD} \right)\frac{\dot \alpha}{\alpha} \nonumber \\
 &\equiv&  H_S(1+S) \frac{\dot h}{h}
 + H_\alpha\frac{\dot \alpha}{\alpha}.
\end{eqnarray}
The last hypothesis assumes that the variations of $h$ and $\alpha$ are
related by Eq.~(\ref{h-alpha}) so that
\begin{eqnarray}\label{nufinal4}
 \frac{\dot\nu_{AB}}{\nu_{AB}}& =& \left[\frac12\left(\hat\lambda_{h}+\frac29  \hat\lambda_{\scriptscriptstyle\rm QCD} \right)(1+S) 
 +\left( \hat\lambda_{\alpha} + R  \hat\lambda_{\scriptscriptstyle\rm QCD} \right)\right]\frac{\dot \alpha}{\alpha} \nonumber \\
   &\equiv& (\frac{1}{2}H_S(1+S) + H_\alpha)\frac{\dot \alpha}{\alpha} \equiv C_\alpha(R,S) \frac{\dot \alpha}{\alpha}.
\end{eqnarray}
The two last equations define the parameter $C_\alpha(R,S)$.

The forms~(\ref{nufinal2}-\ref{nufinal4}) imply increasing assumptions on the unification mechanisms and
are thus becoming more and more model-dependent with the advantage of reducing the number
of fundamental constants, hence allowing one to draw sharper constraints from the same experimental
data.

The coefficients introduced above can be easily calculated from Table~\ref{tab:coef1} for the
coefficients $\lambda$, Tables~\ref{tab:NQM approach results table proton}
or~\ref{tab:chiPT approaches for proton} or Eq.~(\ref{lattice result eq2}) 
for the coefficients $\kappa_i$, Eq.~(\ref{fTqg result}) or
Eq.~(\ref{fTqg isospin eq 2}) for the coefficients $f_{T_i}$, and Eq.~(\ref{coefficients of b varying}) for the coefficients $\gamma_i$. As an example, we provide
the value of the coefficients  $C_\alpha$
assuming $S = 160$ and $ R = 30$ for the 9 models considered in this
article. It is important to stress that this coefficient is almost always larger than one
and typically of order 5 -- 30 in absolute value. 

We can check that the effect of varying the nuclear radius is indeed much smaller than varying the other parameters. This effect can be included by adding a term $\epsilon_r (\dot {\hat{m}} / \hat{m} - {\dot \Lambda_{\scriptscriptstyle\rm QCD} / \Lambda_{\scriptscriptstyle\rm QCD}})$ to Eq.~(\ref{decomposition2}). Using the values listed in Table IV of~\cite{BFK2011}, we have $\epsilon_r = -0.004$ for the Cs-Rb clock system, while $\epsilon_r = -0.007$ for the other five clock systems involving Cs. These amount to an adjustment of  $-0.13$ in the numbers in the first column of Table~\ref{tab-calpha-nqm}, and $-0.23$ in the other five columns. 

\begin{table}[htb!]
\caption{The coefficient $C_\alpha$
assuming $S = 160$ and $ R = 30$ for each of the models for the nucleon magnetic moment
and for the various combinations of clocks discussed in this article.}
\label{tab-calpha-nqm}
\small
\begin{center}
\begin{tabular}{|l|cccccc|}
\hline
 & Cs-Rb & H-Cs & Hg-Cs&  Yb-Cs&  Sr-Cs &SF$_6$-Cs \\
\hline\hline
A &  $-16.53$ &$13.86$ & $17.06$ & $12.96$  & $13.80$    & $4.56$  \\
B1 & $-2.26$  & $20.16$ & $23.36$ & $19.26$ & $20.10$ & $10.85$ \\
B2 &  $-6.79$ & $18.16$ & $21.36$ & $17.26$ & $18.10$ & $8.85$ \\
B3 & $-5.29$  & $18.82$ & $22.02$ & $17.92$ & $18.76$ & $9.51$ \\
C & $-13.37$  & $15.26$ & $18.46$ & $14.36$ & $15.20$ & $5.95$ \\
\hline
HBw/oD & $19.27$  & $29.33$ & $32.53$ & $28.43$ & $29.27$    & $20.22$  \\
HBwD & $-8.57$ & $17.01$ & $20.21$ & $16.11$ & $16.95$    & $7.89$ \\
EOMS& $0.49$  & $20.97$ & $24.17$ & $20.07$ & $20.91$    & $11.86$ \\
\hline
$\chi$PT+QCD& $1.20$  & $21.29$ & $24.49$ & $20.39$ & $21.23$    & $12.18$ \\
\hline
\end{tabular}
\end{center}
\normalsize
\end{table}

\subsection{Single Experiment Constraints}
We are now in a position to combine our results for the dependence of 
the nucleon $g$-factor on fundamental parameters 
with the limits imposed from atomic clock measurements. 
For each experiment, we can derive a limit on the variation of the fine structure constant
under a number of sets of assumptions.

For example, assuming first that the only dependence of $\nu_{AB}$ on 
$\alpha$ is related to the coefficient $\lambda_\alpha$ (i.e., we assume
that $g_{\rm p}$, $g_{\rm n}$, $b$ and $\mu$ remain constant), we can use Table~\ref{tab:coef1}
to extract a limit on $\dot \alpha/\alpha$ for each experiment from
\be
\frac{\dot\alpha}{\alpha} = \frac{1}{\lambda_\alpha} \frac{\dot \nu_{AB}}{\nu_{AB}} .
\ee
In contrast, when we take into account the contributions from 
coupled variations, and we assume the relation between $\dot \nu_{AB}/\nu_{AB}$ and
$\dot \alpha/\alpha$ given by Eq.~(\ref{nufinal4})
we obtain simply
\be
\frac{\dot\alpha}{\alpha} = \frac{1}{C_\alpha} \frac{\dot \nu_{AB}}{\nu_{AB}} .
\ee
Thus, the improvement in the limit from each individual experiment
due to the theoretical assumption of coupled variations is given by
$C_\alpha/\lambda_\alpha$. These factors are tabulated in Table~\ref{enhance}
for each experiment and model for $g_{\rm {p,n}}$.

\begin{table}[htb!]
\caption{The enhancement factor $C_\alpha/\lambda_\alpha$
assuming $S = 160$ and $ R = 30$ for each of the models for the nucleon magnetic moment
and for the various combinations of clocks discussed in this article.}
\label{enhance}
\small
\begin{center}
\begin{tabular}{|l|cccccc|}
\hline
 & Cs-Rb & H-Cs & Hg-Cs&  Yb-Cs&  Sr-Cs &SF$_6$-Cs \\
\hline\hline
A &  $-33.73$ & $4.90$ & $2.83$ & $6.72$  & $4.98$    & $1.61$  \\
B1 & $-4.61$  & $7.12$ & $3.87$ & $9.98$ & $7.26$    & $3.83$ \\
B2 &  $-13.86$ & $6.42$ & $3.54$ & $8.94$ & $6.53$     & $3.13$ \\
B3 & $-10.80$  & $6.65$ & $3.65$ & $9.28$ & $6.77$    & $3.36$ \\
C & $-27.28$  & $5.39$ & $3.06$ & $7.44$ & $5.49$    & $2.10$ \\
\hline
HBw/oD & $39.32$  & $10.36$ & $5.39$ & $14.73$ & $10.57$    & $7.14$  \\
HBwD & $-17.48$ & $6.01$ & $3.35$ & $8.34$ & $6.12$    & $2.79$ \\
EOMS& $1.00$  & $7.41$ & $4.01$ & $10.40$ & $7.55$    & $4.19$ \\
\hline
$\chi$PT+QCD& $2.45$  & $7.52$ & $4.06$ & $10.56$ & $7.66$    & $4.30$ \\
\hline
\end{tabular}
\end{center}
\normalsize
\end{table}

As one can see, there is a strong model-dependence on the resulting limits on 
$\dot\alpha/\alpha$.  Overall the enhancements range from $\sim 1$ to $\sim 10$.
For example, let us consider the case of the Cs-Rb atomic clock system.
Ignoring the variations in all other constants,
this clock would yield a result
\be
\frac{\dot\alpha}{\alpha} = (1.02 \pm 10.82) \times 10^{-16} {\rm yr}^{-1} .
\ee
In contrast, coupled variations, according to the factors in Table~\ref{enhance},
improve this result by as much as a factor of $39.32$ using the HBw/oD model for $g_{\rm {p,n}}$,
yielding 
\be
\frac{\dot\alpha}{\alpha} = (0.03 \pm 0.28) \times 10^{-16} {\rm yr}^{-1} .
\ee
Cases A and C also make substantial improvements in the limit for the Cs-Rb clock system.
On the other hand, there is no gain for case EOMS, or even a {\it weaker} limit if the nuclear radius effect is taken into account.

\subsection{Combined Experimental Constraints}

While the results of individual experiments can be substantially improved
by coupled variations,  two clock systems (Dy and Hg-Al) are independent 
of any assumption on unification and lead to model-independent limits on $\alpha$.
We next combine the available results to obtain a single limit on $\alpha$ for each choice 
of model for $g_{\rm p,n}$.

Each of the eight experimental results used in this article can be written as
\begin{equation}
  \frac{\dd}{\dd t}\ln \nu_{AB} = \eta_{AB} \pm\delta_{AB},
\end{equation}
listed in Table~\ref{tab:coef1}.
From a theoretical point of view, the expression for $\nu_{AB}$ depends
on a set of constants, ${\bf x}$, chosen
as being independent and on our hypothesis on unification schemes.
If we assume $d \ln \nu_{AB}({\bf x})/dt -\eta_{AB}$ to be Gaussian distributed and
all the experiments to be uncorrelated, then the best-fit
for the set of constants ${\bf x}$ is obtained by maximizing
the likelihood, or equivalently by minimizing
\begin{equation}
\chi^2({\bf x}) = \sum_{AB} \frac{\left[\frac{\dot\nu_{AB}}{\nu_{AB}}({\bf x})-\eta_{AB}\right]^2}{\delta_{AB}^2}.
\end{equation}
The   68.27\%, 95\%, and 99\% confidence level
(i.e., $1\sigma$, $\sim 2\sigma$ and $\sim 3\sigma$) constraints are then obtained by
$\Delta\chi^2 =(1,3.84,6.63)$ if dim$({\bf x})=1$ and $\Delta \chi^2 =(2.30,5.99,9.21)$ if dim$({\bf x})=2$.

\subsubsection{Constraints on the QED parameters}

\begin{figure}[ht!]
\begin{center}
\epsfig{file=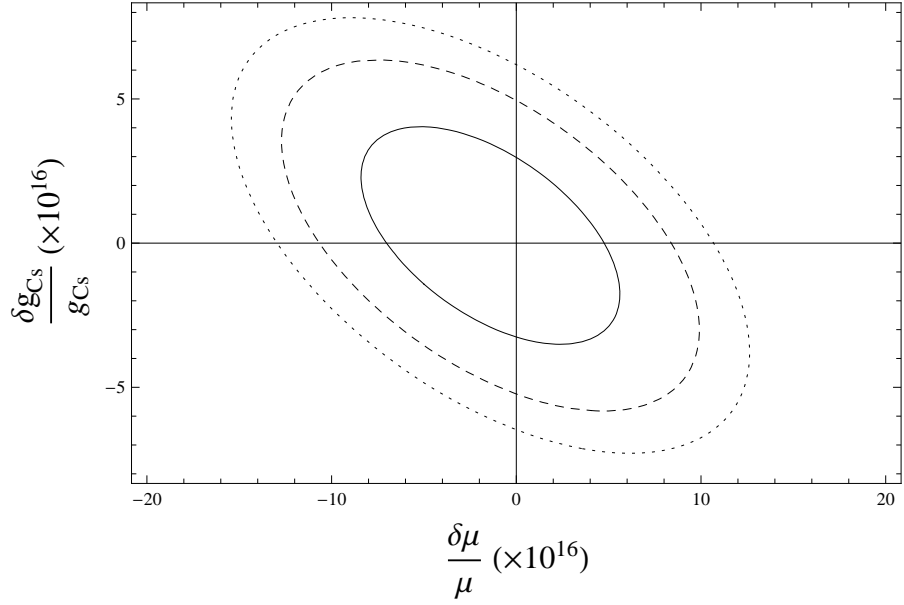,height=8cm}
\end{center}
\caption{Constraints on the variation of parameters $\lbrace g_{\rm Cs},\mu\rbrace$ assumed
to be independent once the constraint from the variation of $\alpha$ from the Hg-Al clock
is taken into account. Solid, dashed and dotted contours correspond to 68.27\%, 95\%
and 99\% C.L.}
\label{fig:contrainteQED}
\end{figure}

Let us start by assuming that $\lbrace g_{\rm p}, g_{\rm n}, b, \mu,\alpha\rbrace$ are independent parameters.
One can use the Hg-Al clock to constrain the variation of $\alpha$ and then use the
six clock combinations that depend on the five parameters to set a constraint on
$\lbrace g_{\rm p}, g_{\rm n}, b, \mu\rbrace$. However, from Eq.~(\ref{preferred method formula}), we note that the ratio of the coefficients of 
$\delta g_{\rm n} / g_{\rm n}$ and $\delta b / b$ is $g_{\rm n} / (g_{\rm n} - g_{\rm p} + 1)$, which is independent of the clock systems we are considering. Also, from Table~\ref{tab:coef1}, we note that the value of $\lambda_{g_{\rm p}} / \lambda_{g_{\rm n}}$ for the Cs-Rb clock is very close to that of the other five clock combinations. Therefore, for the purpose of constraining the QED parameters,  $g_{\rm p}$, $g_{\rm n}$ and $b$ are not independent, and we can only constrain their combination, namely, $g_{\rm Cs}$. The combined constraint on $g_{\rm Cs}$ and $\mu$ is depicted on Fig.~\ref{fig:contrainteQED}. Note that if a different method in the calculation of $g$-factors of \super{87}Rb and \super{133}Cs, and/or other clock systems, are used, such that the ratio $\lambda_{g_{\rm p}} : \lambda_{g_{\rm n}} : \lambda_b$ is not the same for different clock combinations, then $g_{\rm p}$, $g_{\rm n}$ and $b$ can be taken as independent parameters. 

As we know from our analysis, such a hypothesis is not correct since the variations are
expected to be correlated but this shows the result one would have derived without
any knowledge on QCD.

\subsubsection{Constraint on $\alpha$}

As in the previous subsection, we can consider first the constraint 
obtained using the form~(\ref{nufinal4}) that depends
on $\delta\alpha/\alpha$ alone. Minimizing $\chi^2$ for a single variable is equivalent to taking the weighted
mean of $\eta_{AB}/\lambda_\alpha$ with an uncertainty $\delta_{AB}/\lambda_\alpha$.
This result can be compared with that assuming coupled variations using
the coefficients $C_\alpha$, given in Table~\ref{tab-calpha-nqm}. In this case, the weighted mean replaces
$\lambda_\alpha$ with $C_\alpha$.

In order to determine the effect of coupled variations, we compare the constraints
arising from the combination of the eight experiments to the one obtained
from the combination of 6 clocks (that is neglecting the Dy and Hg-Al clocks). The results
are presented in Table~\ref{tab-alpha-nqm} and shall be compared to the
same analysis assuming that only $\alpha$ is varying (i.e., keeping $g_{\rm p}$, $g_{\rm n}$, $b$ and $\mu$ constant).  We find
\begin{equation}
 \frac{\dot\alpha}{\alpha} = -(2.14\pm2.30)\times 10^{-17}\,{\rm yr}^{-1}
\end{equation}
for the combination of the 8 experiments and
\begin{equation}
 \frac{\dot\alpha}{\alpha} = -\left(5.24\pm6.40 \right)\times 10^{-17}\,{\rm yr}^{-1}
\end{equation}
for the combination of the 6 experiments. We also remind the reader that
the Hg-Al experiment alone set the constraint
\begin{equation}
 \frac{\dot\alpha}{\alpha} = -\left(1.65\pm2.46\right)\times 10^{-17}\,{\rm yr}^{-1},
 \label{hgalresult}
\end{equation}
which shows that there is little gain in combining the 8 experiments compared
to this experiment alone.

When $g_{\rm p}$, $g_{\rm n}$, $b$ and $\mu$ are allowed to vary in the combination of the 6 clocks,
there is a gain of a factor of order 4 so that the constraint obtained from the
combination of these 6 clocks assuming unification becomes as strong as 
the constraint obtained from Hg-Al alone. When combining the 8 experiments,
the gain is less than a factor of 2, due to the fact that the limit arises mostly from
the Hg-Al experiment which does not depend on $g_{\rm p,n}$.
These results are summarized in Table~\ref{tab-alpha-nqm} and each result
can be compared to the single Hg-Al result given in Eq.~(\ref{hgalresult}).

\begin{table}[htb!]
\caption{Constraints on the variation of $\alpha$
assuming the unification relation~(\ref{nufinal4}) and the values of $C_\alpha$
for $S = 160$ and $R = 30$. We compare  the constraints obtained from
the combination of the 8 clocks and  the constraints obtained from the 6 
clocks (i.e. without Dy and Hg-Al). All numbers are in yr$^{-1}$.
}
\label{tab-alpha-nqm}
\small
\begin{center}
\begin{tabular}{|c|cc|}
\hline
 Model & 8 clocks &  6 clocks   \\
\hline\hline
 A   & $(-1.32 \pm1.46)\,\times10^{-17}$ & $ (-1.12\pm1.81)\,\times10^{-17}$\\
 B1 & $ (-1.25\pm1.34)\,\times10^{-17}$ & $ (-1.07\pm1.60)\,\times10^{-17}$\\
 B2& $ (-1.32\pm1.40)\,\times10^{-17}$ & $(-1.15\pm1.71)\,\times10^{-17}$\\
 B3& $ (-1.30\pm1.38)\,\times10^{-17}$ & $ (-1.13\pm1.67)\,\times10^{-17}$\\
 C& $(-1.35\pm1.46)\,\times10^{-17}$    & $(-1.17\pm1.81)\,\times10^{-17}$\\
 \hline
 HBw/oD& $ (-0.76\pm0.97)\,\times10^{-17}$ & $ (-0.60\pm1.06)\,\times10^{-17}$\\
 HBwD& $ (-1.36\pm1.44)\,\times10^{-17}$ & $ (-1.19\pm1.78)\,\times10^{-17}$\\
 EOMS& $ (-1.21\pm1.31)\,\times10^{-17}$ & $ (-1.02\pm1.54)\,\times10^{-17}$\\
 \hline
 $\chi$PT+QCD& $ (-1.19\pm1.30)\,\times10^{-17}$ & $(-1.00\pm1.52)\,\times10^{-17}$\\
\hline
\end{tabular}
\end{center}
\normalsize
\end{table}

\subsubsection{Constraint on $hv$}

As a second application, we can use the constraint~(\ref{clock-bound1}) 
arising from the Hg-Al clock to obtain a bound on the time variation of
$\alpha$ that is independent of the other constants
and then use the 6 other clocks to set a constraint on the 
combination of parameters $hv$, assuming
the form~(\ref{nufinal2}) to set a constraint on $\delta hv/hv$ alone. 
This requires the knowledge of the
coefficients $H_S$ and $H_\alpha$ and we assume that $R = 30$, but it does not
depend on the coefficient $S$.

The constraints for each model are summarized on Table~\ref{tab-hv-nqm}.
It ranges between $\left|\frac{(hv)^.}{hv}\right|<20.43\times10^{-16}$~yr$^{-1}$
and $\left|\frac{(hv)^.}{hv}\right|<13.52\times10^{-16}$~yr$^{-1}$, respectively for models A and 
HBw/oD and it turns out that the model-dependence for this constraint
is mild.

\begin{table}[htb!]
\caption{Constraints on the variation of $hv$
once the variation of $\alpha$ alone is constrained from the Hg-Al clock.
It assumes the unification relation~(\ref{nufinal2}). All numbers are in yr$^{-1}$.}
\label{tab-hv-nqm}
\small
\begin{center}
\begin{tabular}{|c|c|}
\hline
 Model & $\frac{(hv)^.}{hv}$  \\
\hline\hline
 A   & $(-9.64\pm10.79)\,\times10^{-16}$\\
 B1 &$(-8.69\pm10.21)\,\times10^{-16}$\\
 B2 &$(-9.27\pm10.52)\,\times10^{-16}$\\
 B3 &$(-9.11\pm10.45)\,\times10^{-16}$\\
 C   &$(-9.65\pm10.71)\,\times10^{-16}$\\
 \hline
 HBw/oD &$(-6.20\pm7.32)\,\times10^{-16}$\\
 HBwD    &$(-9.55\pm10.76)\,\times10^{-16}$\\
 EOMS    &$(-8.35\pm10.01)\,\times10^{-16}$\\
 \hline
 $\chi$PT+QCD &$(-8.22\pm9.91)\,\times10^{-16}$\\
\hline
\end{tabular}
\end{center}
\normalsize
\end{table}

\subsubsection{Constraint on $(\frac{hv}{\Lambda_{\scriptscriptstyle\rm QCD}},\alpha)$}

As a third application, we use the fact that $\hat\lambda_{\scriptscriptstyle\rm QCD}=-\hat\lambda_h$ so that
the form~(\ref{nufinal1red}) allows one to set a constraint on $(hv/\Lambda_{\scriptscriptstyle\rm QCD},\alpha)$
independent of any hypothesis on unification and thus does not
require knowledge of the parameters $R$ and $S$.

\begin{figure}[ht!]
\begin{center}
\epsfig{file=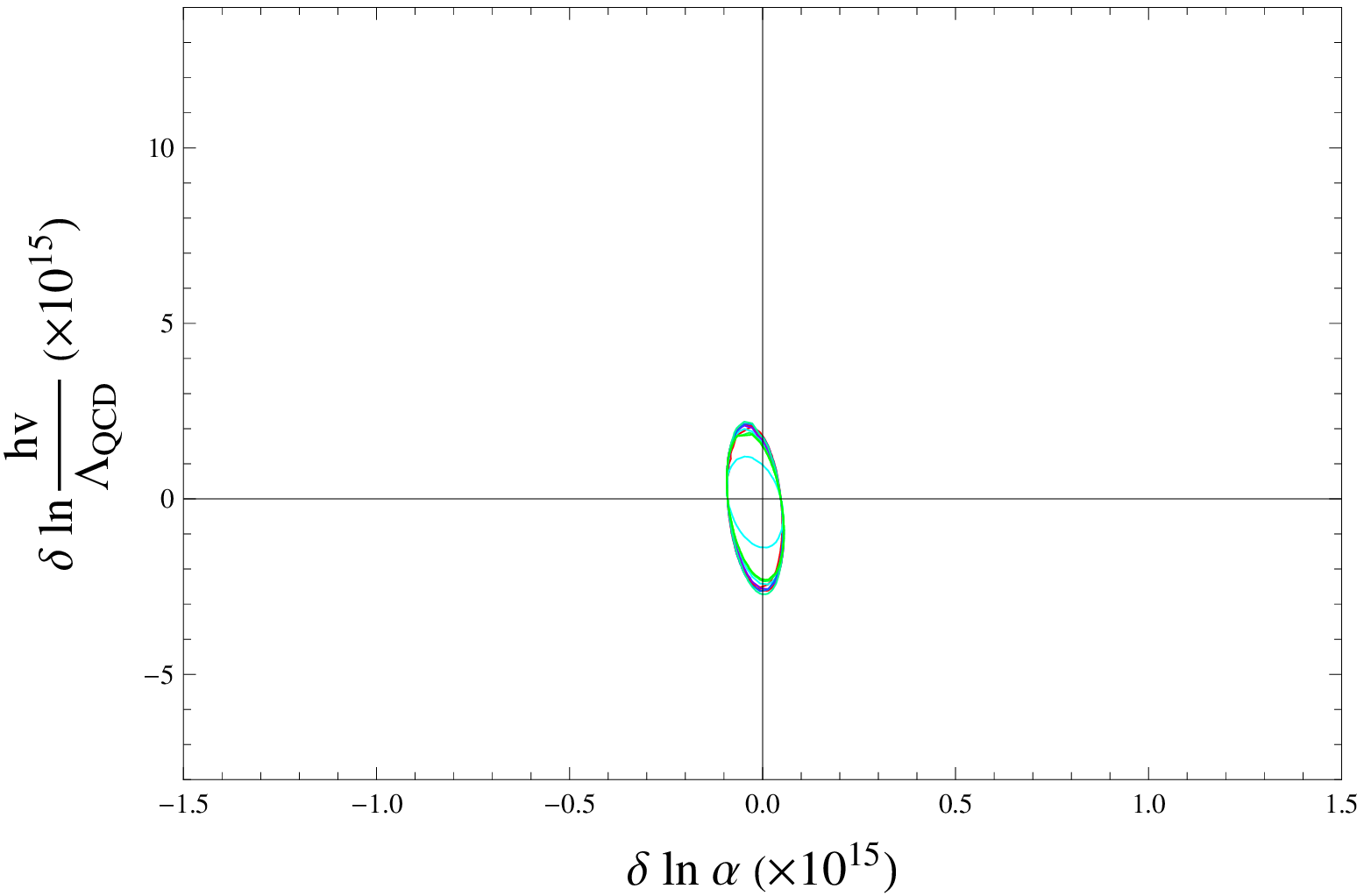,height=5.1cm}
\epsfig{file=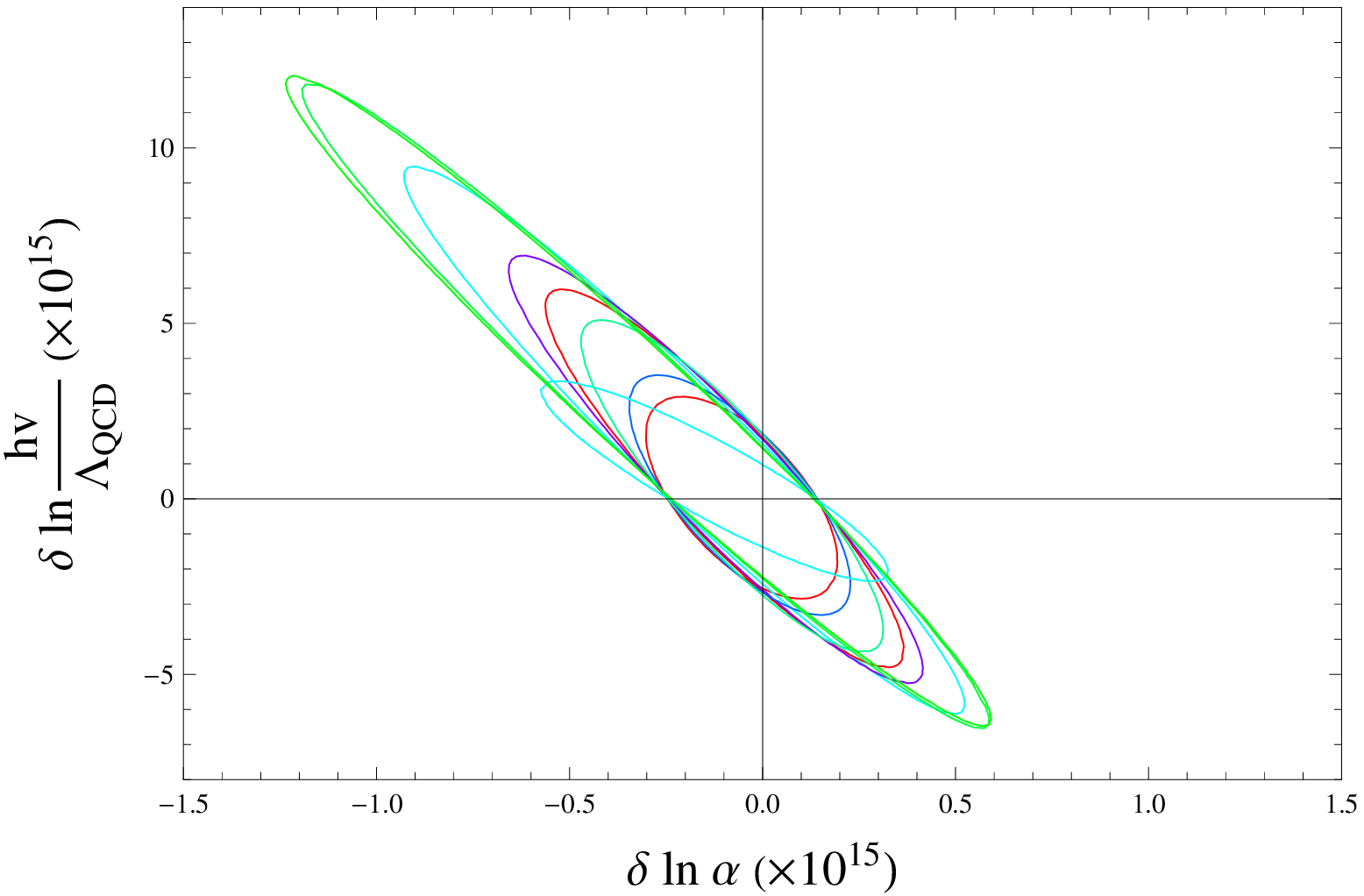,height=5.1cm}
\end{center}
\caption{Comparison of the 99\% C.L. constraints on $(hv/\Lambda_{\scriptscriptstyle\rm QCD},\alpha)$
for the 9 models with 8 clocks (left) and 6 clocks (right)}
\label{fig:contrainteNQM3}
\end{figure}

Figure~\ref{fig:contrainteNQM3} compares the 99\% C.L. constraints obtained from
the combination of 6 and 8 experiments for each model. Again, we see that 
the Hg-Al experiment dominates the collective limit.

\section{Application to astrophysical systems and Discussion}

\subsection{Astrophysical systems}

Several different types of observations of astrophysical systems involving
quasar absorption spectra are subject to a
similar analysis that has been applied to atomic clocks.
Indeed, there are four distinct combinations of physical parameters which depend on $g_{\rm p}$.

\begin{itemize}
\item The comparison of UV heavy element transitions with the hyperfine H\,{\sc i} transition 
allows one to set constraints on
\begin{equation}
 x\equiv\alpha^2 g_{\mathrm{p}} \mu,
\end{equation}
since the  optical transitions are simply proportional to $R_\infty$. It follows that constraints on the time
variation of $x$ can be obtained from high resolution 21~cm spectra
compared to UV lines, e.g., of Si\,{\sc ii}, Fe\,{\sc ii} and/or
Mg\,{\sc ii}. The recent detection of 21~cm and molecular hydrogen absorption lines
in the same damped Lyman-$\alpha$ system at $z_{\mathrm{abs}}=3.174$
towards SDSS J1337+3152 constrains~\cite{x21cm} the variation $x$ to
\begin{equation}
 \Delta x/x=-(1.7\pm1.7)\times10^{-6},\qquad z=3.174.
\end{equation}

\item The comparison of the  H\,{\sc i} 21~cm hyperfine transition 
to the rotational transition frequencies of diatomic molecules 
allows one to set a constraint on
\begin{equation}
y \equiv g_{\mathrm{p}}\alpha^2
\end{equation}
The most recent constraint~\cite{y-murph} relies on the comparison
of two absorption systems determined
both from H\,{\sc i} and molecular absorption. The first is a system at
$z=0.6847$ in the direction of TXS~0218+357 for which the spectra of
CO(1-2), \super{13}CO(1-2), C\super{18}O(1-2), CO(2-3),
HCO\super{+}(1-2) and HCN(1-2) are available. They concluded that
\begin{equation}
 \Delta y/y=(-0.16\pm0.54)\times10^{-5}, \qquad z=0.6847.
\end{equation}
The second system is an absorption system in the direction of
PKS~1413+135 for which the molecular lines of CO(1-2),
HCO\super{+}(1-2) and HCO\super{+}(2-3) have been detected. The analysis
led to
\begin{equation}
 \Delta y/y=(-0.2\pm0.44)\times10^{-5},\qquad z=0.247.
\end{equation}

\item The ground state, ${}^2\Pi_{3/2} J=3/2$, of OH is split into
two levels by $\Lambda$-doubling and each of these doubled levels
is further split into two hyperfine-structure states. Thus, it has
two ``main'' lines ($\Delta F=0$) and two ``satellite'' lines
($\Delta F=1$). Since these four lines arise from two different
physical processes ($\Lambda$-doubling and hyperfine splitting),
they enjoy the same Rydberg dependence but different $g_{\mathrm{p}}$
and $\alpha$ dependences. By comparing the four transitions to the
H\,{\sc i} hyperfine line, one can set a constraint on
\begin{equation}
 F\equiv g_{\mathrm{p}}(\alpha^2/\mu)^{1.57}.
\end{equation}
Using the four 18~cm OH lines from the gravitational lens at
$z\sim0.765$ toward PMN~J0134-0931 and comparing the H\,{\sc i} 21~cm
and OH absorption redshifts of the different components allowed one to set
the constraint~\cite{OH-3}
\begin{equation}
 \Delta F/F=(-0.44\pm0.36\pm1.0_{\text{syst}})\times10^{-5},\qquad
 z=0.765,
\end{equation}
where the second error is due to velocity offsets between OH and
H\,{\sc i} assuming a velocity dispersion of 3~km/s. A similar
analysis~\cite{darling} in a system in the direction of
PKS~1413+135 gave
\begin{equation}
 \Delta F/F=(0.51\pm1.26)\times10^{-5},\qquad
 z=0.2467.
\end{equation}

\item The satellite OH~18~cm lines are conjugate so that the two lines have
the same shape, but with one line in emission and the other in
absorption.  This behavior
has recently been discovered at cosmological distances  and it was
shown~\cite{OH-1} that a comparison between the sum and difference of
satellite line redshifts probes the variation of
\begin{equation}
 G \equiv g_{\mathrm{p}}(\alpha^2/\mu)^{1.85}.
\end{equation} 
From the analysis of a system at
$z\sim0.247$ towards PKS~1413+135, it was concluded~\cite{kanekar1} that
$|\Delta G/G| =(2.2\pm3.8) \times10^{-5}$, while a newer analysis~\cite{kanekar1bis} gave
\begin{equation}
|\Delta G/G| =(-1.18\pm0.46) \times10^{-5}.
\end{equation}
It was also applied to a nearby system~\cite{kanekar2}, Centaurus~A, to give $ |\Delta
G/G| < 1.16\times10^{-5}$ at $z\sim0.0018$. 
\end{itemize}

These constraints are summarized in Table~\ref{tab:astro}.

\begin{table}[htb!]
\caption{Constraints on the variation of different combinations of
$g_{\rm p}$, $\mu$ and $\alpha$ from astrophysical observations.}
\label{tab:astro}
\small
\begin{center}
\begin{tabular}{|l|ccc|cc|}
\hline
Combination & $\lambda_{g_{\rm p}}$ & $\lambda_\mu$ & $\lambda_\alpha$ & Constraints (yr$^{-1}$) & redshift  \\
\hline\hline
 $x=g_{\mathrm{p}} \alpha^2 \mu$ & $1$ &$1$ & $2$ & $-(1.7\pm1.7)\times10^{-6}$ & $3.174$ \\
  $y=g_{\mathrm{p}}\alpha^2$ & $1$ & $0$ & $2$ & $(-0.16\pm0.54)\times10^{-5}$ & $0.6847$ \\
         & & & & $(-0.2\pm0.44)\times10^{-5}$ & $0.247$ \\
   $F=g_{\mathrm{p}}(\alpha^2/\mu)^{1.57}$ & $1$ & $-1.57$ & $3.14$ & $(-0.44\pm0.36\pm1.0_{\text{syst}})\times10^{-5}$ & $0.765$ \\    
   & & & & $(0.51\pm1.26)\times10^{-5}$ & $0.2467$ \\
$G=g_{\mathrm{p}}(\alpha^2/\mu)^{1.85}$ & 1 &$-1.85$ & 3.70 & $(-1.18\pm0.46)\times10^{-5}$& $0.247$ \\
    & & & & $(0\pm1.16)\times10^{-5}$ & $0.0018$ \\   
\hline
\end{tabular}
\end{center}
\normalsize
\end{table}

\subsection{Astrophysical constraints}

In contrast to our analysis of atomic clocks, we cannot combine the astrophysical
observations because they have been obtained from different
systems at different redshifts and at different spatial locations.
However, as we have done previously (but without the $g_{\rm n}$ and $b$ terms), we show in 
Table~\ref{tab:factorastro} the enhancement factor for
the analysis of the 4 types of combinations of absorption spectra. We 
emphasize that the enhancement factor is always larger than unity
(except for $y$ in the EOMS and $\chi$PT+QCD models).  As last example of the power of coupled
variations, Table~\ref{tab:astrocte}
compares the constraints on the variation of $\alpha$ that can be
obtained under the assumption that $g_{\rm p}$ and $\mu$ are
constant with the assumption of coupled variations based on unification.
As one can see, in many cases the limits are improved by an order of magnitude.

\begin{table}[htb!]
\caption{Value of the parameter $C_\alpha$ for the 4 combinations
of constants that can be constrained by astrophysical observations,
assuming $R=30$ and $S=160$ (left) and value of the enhancement
factor $C_\alpha /\lambda_\alpha$ (right). }
\label{tab:factorastro}
\small
\begin{center}
\begin{tabular}{|l|cccc|}
\hline
 & $x$ & $y$ & $F$ &  $G$ \\
\hline\hline
A & $34.10$ & $15.49$ & $-12.60$ & $-17.25$ \\
B1  & $20.72$ & $2.10$ & $-25.98$ & $-30.64$ \\
B2 & $24.96$  & $6.35$ & $-21.73$ & $-26.39$ \\
B3  & $23.52$ & $4.91$ & $-23.18$ & $-27.83$ \\
C &  $31.21$  & $12.59$ & $-15.49$ & $-20.14$ \\
\hline
HBw/oD  & $2.46$ & $-15.76$ & $-43.23$ & $-47.77$ \\
HBwD  &  $27.00$ & $8.78$ & $-18.69$ & $-23.23$ \\
EOMS &  $17.63$ & $-0.60$ & $-28.07$ & $-32.61$ \\
\hline
$\chi$PT+QCD & $16.96$ & $-1.26$ & $-28.73$ & $-33.27$ \\
\hline
\end{tabular}
\begin{tabular}{|l|cccc|}
\hline
 & $x$ & $y$ & $F$ &  $G$ \\
\hline\hline
A   &  $17.05$  & $7.74$ & $-4.01$ & $-4.66$ \\
B1  & $10.36$  & $1.05$ & $-8.27$ & $-8.28$ \\
B2  & $12.48$  & $3.18$ & $-6.92$ & $-7.13$ \\
B3  &  $11.76$ & $2.45$ & $-7.38$ & $-7.52$ \\
C   &  $15.60$ & $6.30$ & $-4.93$ & $-5.44$ \\
\hline
HBw/oD  & $1.23$ & $-7.88$ & $-13.77$ & $-12.91$ \\
HBwD  & $13.50$ & $4.39$ & $-5.95$ & $-6.28$ \\
EOMS & $8.81$ & $-0.30$ & $-8.94$ & $-8.81$ \\
\hline
$\chi$PT+QCD & $8.48$ &$-0.63$ & $-9.15$ & $-8.99$ \\
\hline
\end{tabular}
\end{center}
\normalsize
\end{table}

\begin{table}[htb!]
\caption{Comparison of the constraints obtained from astrophysical systems
with and without assumption on unification for model A.}
\label{tab:astrocte}
\small
\begin{center}
\begin{tabular}{|l|ccc|}
\hline
Combination & independent (yr$^{-1}$) & correlated (yr$^{-1}$) & redshift  \\
\hline\hline
 $x$ & $(-8.50\pm8.50)\times10^{-7}$ & $(-4.98\pm4.98)\times10^{-8}$ & 3.174 \\
  $y$ & $(-0.8\pm2.7)\times10^{-6}$ & $(-1.03\pm3.49)\times10^{-7}$ & $0.6847$ \\
        & $(-1.0\pm2.2)\times10^{-6}$ & $(-1.29\pm2.84)\times10^{-7}$ & $0.247$ \\
   $F$ & $(-1.40\pm3.38)\times10^{-6}$ & $(3.49\pm8.44)\times10^{-7}$& $0.765$ \\    
   & $(1.62\pm4.01)\times10^{-6}$& $(-0.40\pm1.00)\times10^{-6}$ & $0.2467$ \\
$G$ & $(-3.19\pm1.24)\times10^{-6}$ & $(6.84\pm 2.67)\times10^{-7}$& $0.247$ \\
   & $(0\pm3.14)\times10^{-6}$ & $(0\pm6.73)\times10^{-7}$& $0.0018$ \\   
\hline
\end{tabular}
\end{center}
\normalsize
\end{table}

\subsection{Discussion}

In this article, we have discussed the effect of a correlated variation of fundamental constants,
focusing on the gyromagnetic factors $g_{\rm p}$ and $g_{\rm n}$. These parameters are particularly
important to interpret electromagnetic spectra, and thus to derive constraints
on the variation of fundamental constants from atomic clock experiments
and from quasar absorption spectra.
As discussed, there is an important model-dependence in the computation
of the gyromagnetic factors in terms of the quark masses and QCD scale.

When applied to the interpretation of atomic clock experiments, we have shown that
in general the constraints on the variation of $\alpha$ are sharper than that under
the assumption that $g_{\rm p}$, $g_{\rm n}$, $b$ and $\mu$ are constant, but this is not a systematic
conclusion as we have exhibited models in which the variation of $\alpha$ stays the same or
is even weaker due to cancellations in the sensitivity to $\alpha$. The constraints
on the variation of $\alpha$ should then be taken with care. In many cases, they may
be stronger than reported, but they may
be weaker as well.  This points to the need to better understand
the fundamental physics needed to calculate baryon magnetic moments.
Any limit which depends on $g_{\rm p,n}$ will be subject to the type of uncertainties discussed here.

Fortunately, the tightest constraint arises
from the Hg-Al clock experiments, that does not depend on $g_{\rm p}$, $g_{\rm n}$, $b$ or $\mu$. As a consequence, we have been able to independently
set a bound on the variation of $hv$ from the combination of the
other experiments. While this bound is still model-dependent, we have
shown that it is always smaller than
\begin{equation}
\left|\frac{(hv)^.}{hv}\right|<2.0\times10^{-15}\,{\rm yr}^{-1}
\end{equation}
for the models we have considered in this article.

Our analysis also applies to astrophysical system and to quasar absorption spectra.
We have shown that the enhancement factor is almost always larger than unity.


\section*{Acknowledgments}
We would like to thank X. Cui, J. Ellis, M. Peskin, M. Srednicki, A. Vainshtein, and M. Voloshin for helpful 
discussions.
The work of FL and KAO  was supported in part by
DOE grant
DE-FG02-94ER-40823 at the University of Minnesota. JPU was partially supported by the ANR/Thales.



\begin{thebibliography}{99}

\bibitem{Bek}
  P. Jordan, 
  Die Naturwissenschaften {\bf25}, 513 (1937);\\
  M. Fierz,
  Helv. Phys. Acta {\bf29}, 128 (1956);\\
  C. Brans, and R. Dicke, 
  Phys. Rev. {\bf124}, 925 (1961);\\
  R.H. Dicke, 
  in DeWitt, C.M. and DeWitt, B.S., eds., Relativity, Groups and Topology. Relativit\'e, Groupes et Topologie, 
  Lectures   delivered at Les Houches during the 1963 session of the Summer School of Theoretical Physics, 
  University  of Grenoble, pp. 165Ð313, (Gordon and Breach, New York; London, 1964);\\
  J.~D.~Bekenstein,
  Phys.\ Rev.\ D {\bf 25}, 1527 (1982).

\bibitem{webb}
 J.~K.~Webb, V.~V.~Flambaum, C.~W.~Churchill, M.~J.~Drinkwater, \etal,
 Phys.\ Rev.\ Lett.\  {\bf 82}, 884 (1999),
 \url{arXiv:astro-ph/9803165};\\
 M.~T.~Murphy {\it et al.},
 Mon.\ Not.\ Roy.\ Astron.\ Soc.\  {\bf 327}, 1208 (2001),
 \url{arXiv:astro-ph/0012419};\\
 J.~K.~Webb, \etal,
 Phys.\ Rev.\ Lett.\  {\bf 87}, 091301 (2001),
 \url{arXiv:astro-ph/0012539};\\
 M.~T.~Murphy, J.~K.~Webb, V.~V.~Flambaum, C.~W.~Churchill, \etal,
 Mon.\ Not.\ Roy.\ Astron.\ Soc.\  {\bf 327}, 1223 (2001),
 \url{arXiv:astro-ph/0012420}.

\bibitem{murphy3}
  M.~T.~Murphy, J.~K.~Webb and V.~V.~Flambaum,
  Mon.\ Not.\ Roy.\ Astron.\ Soc.\  {\bf 345}, 609 (2003),
  \url{arXiv:astro-ph/0306483}.

\bibitem{chand}
  H.~Chand, R.~Srianand, P.~Petitjean and B.~Aracil,
  Astron.\ Astrophys.\  {\bf 417}, 853 (2004),
  \url{arXiv:astro-ph/0401094};\\
  R.~Srianand, H.~Chand, P.~Petitjean and B.~Aracil,
  Phys.\ Rev.\ Lett.\  {\bf 92}, 121302 (2004),
  \url{arXiv:astro-ph/0402177}.

\bibitem{quast}
  R.~Quast, D.~Reimers and S.~A.~Levshakov,
  Astron.\ Astrophys.\  {\bf 415}, L7 (2004),
  \url{arXiv:astro-ph/0311280}.

\bibitem{murphy07}
  M.~T.~Murphy, J.~K.~Webb and V.~V.~Flambaum,
  Phys.\ Rev.\ Lett.\  {\bf 99}, 239001 (2007),
  \url{arXiv:0708.3677 [astro-ph]}.

\bibitem{chand3}
  R.~Srianand, H.~Chand, P.~Petitjean and B.~Aracil,
  Phys.\ Rev.\ Lett.\  {\bf 99}, 239002 (2007).
  
\bibitem{ekow}
  J.~R.~Ellis, S.~Kalara, K.~A.~Olive and C.~Wetterich,
  Phys.\ Lett.\  B {\bf 228}, 264 (1989).
  
\bibitem{co}
  B.~A.~Campbell and K.~A.~Olive,
  Phys.\ Lett.\ B {\bf 345}, 429 (1995),
  \url{arXiv:hep-ph/9411272}.

\bibitem{lang}
  P.~Langacker, G.~Segre and M.~J.~Strassler,
  Phys.\ Lett.\ B {\bf 528}, 121 (2002),
  \url{arXiv:hep-ph/0112233};\\
  T.~Dent and M.~Fairbairn,
  Nucl.\ Phys.\  B {\bf 653}, 256 (2003),
  \url{arXiv:hep-ph/0112279};
  X.~Calmet and H.~Fritzsch,
  Eur.\ Phys.\ J.\  C {\bf 24}, 639 (2002),
  \url{arXiv:hep-ph/0112110};\\
  X.~Calmet and H.~Fritzsch,
  Phys.\ Lett.\  B {\bf 540}, 173 (2002),
  \url{arXiv:hep-ph/0204258};\\
  T.~Damour, F.~Piazza and G.~Veneziano,
  Phys.\ Rev.\ Lett.\  {\bf 89}, 081601 (2002),
  \url{arXiv:gr-qc/0204094};\\
  T.~Damour, F.~Piazza and G.~Veneziano,
  Phys.\ Rev.\  D {\bf 66}, 046007 (2002),
  \url{arXiv:hep-th/0205111}.

\bibitem{dine}
 M.~Dine, Y.~Nir, G.~Raz and T.~Volansky,
  Phys.\ Rev.\  D {\bf 67}, 015009 (2003),
  \url{arXiv:hep-ph/0209134}.
  
\bibitem{eos}
  J.~R.~Ellis, K.~A.~Olive and Y.~Santoso,
  New J.\ Phys.\  {\bf 4}, 32 (2002),
  \url{arXiv:hep-ph/0202110}.
  
\bibitem{ds}
   V.V. Dixit and M. Sher, Phys. Rev. D {\bf37} (1988) 1097.
   
\bibitem{uzan}
  J.-P.~Uzan,
  Rev.\ Mod.\ Phys.\  {\bf 75}, 403 (2003),
  \url{arXiv:hep-ph/0205340};\\
  J.-P. Uzan,
  AIP Conf. Proc. {\bf736}, 3 (2005), \url{astro-ph/0409424};\\
 J.-P. Uzan,
 Space Sci. Rev. {\bf148}, 249 (2010), \url{arXiv:0907.3081};\\
 G.F.R. Ellis and J.-P. Uzan,
 Am. J. Phys. {\bf73}, 240 (2005), \url{gr-qc/0305099}.

\bibitem{uzan2}
  J.~P.~Uzan,
  Living Rev.\ Rel.\  {\bf 14}, 2 (2011),
  \url{arXiv:1009.5514 [astro-ph.CO]}.

\bibitem{ichikawa}
  K.~Ichikawa and M.~Kawasaki,
  Phys.\ Rev.\  D {\bf 65}, 123511 (2002),
  \url{arXiv:hep-ph/0203006}.
  
\bibitem{wett}
  C.~M.~Muller, G.~Schafer and C.~Wetterich,
  Phys.\ Rev.\  D {\bf 70}, 083504 (2004),
  \url{arXiv:astro-ph/0405373};\\
  T.~Dent, S.~Stern and C.~Wetterich,
  Phys.\ Rev.\  D {\bf 76}, 063513 (2007),
  \url{arXiv:0705.0696 [astro-ph]}.

\bibitem{cnouv}
 A.~Coc, N.~J.~Nunes, K.~A.~Olive, J.~P.~Uzan, \etal,
  Phys.\ Rev.\  D {\bf 76}, 023511 (2007),
  \url{arXiv:astro-ph/0610733}.
  
\bibitem{df}
  V.~V.~Flambaum and E.~V.~Shuryak,
  Phys.\ Rev.\  D {\bf 65}, 103503 (2002),
  \url{arXiv:hep-ph/0201303};\\
  V.~F.~Dmitriev and V.~V.~Flambaum,
  Phys.\ Rev.\  D {\bf 67}, 063513 (2003),
  \url{arXiv:astro-ph/0209409};\\
  V.~V.~Flambaum and E.~V.~Shuryak,
  Phys.\ Rev.\  D {\bf 67}, 083507 (2003),
  \url{arXiv:hep-ph/0212403};\\
  V.~F.~Dmitriev, V.~V.~Flambaum and J.~K.~Webb,
  Phys.\ Rev.\  D {\bf 69}, 063506 (2004),
  \url{arXiv:astro-ph/0310892};\\
  J.~C.~Berengut, V.~V.~Flambaum and V.~F.~Dmitriev,
  Phys.\ Lett.\  B {\bf 683}, 114 (2010),
  \url{arXiv:0907.2288 [nucl-th]}.
  
  \bibitem{landau}
  S.~J.~Landau, M.~E.~Mosquera, C.~G.~Scoccola and H.~Vucetich,
  Phys.\ Rev.\  D {\bf 78}, 083527 (2008)
  [arXiv:0809.2033 [astro-ph]].
  
  \bibitem{grant}
  M.~K.~Cheoun, T.~Kajino, M.~Kusakabe and G.~J.~Mathews,
  arXiv:1104.5547 [astro-ph.CO].

\bibitem{opqccv}
K.~A.~Olive, M.~Pospelov, Y.~Z.~Qian, A.~Coc, \etal,
 Phys.\ Rev.\ D {\bf 66}, 045022 (2002),
 \url{arXiv:hep-ph/0205269}.

\bibitem{opqmvcc}
 K.~A.~Olive, M.~Pospelov, Y.~Z.~Qian, G.~Manhes, \etal,
  Phys.\ Rev.\  D {\bf 69}, 027701 (2004),
  \url{arXiv:astro-ph/0309252}.
  
  \bibitem{wett2}
  T.~Dent, S.~Stern and C.~Wetterich,
  Phys.\ Rev.\  D {\bf 78}, 103518 (2008)
  [arXiv:0808.0702 [hep-ph]];
  T.~Dent, S.~Stern and C.~Wetterich,
  Phys.\ Rev.\  D {\bf 79}, 083533 (2009)
  [arXiv:0812.4130 [hep-ph]].
  
  \bibitem{cmb}
  M.~Nakashima, K.~Ichikawa, R.~Nagata and J.~Yokoyama,
  JCAP {\bf 1001}, 030 (2010)
  [arXiv:0910.0742 [astro-ph.CO]];\\
  C.~J.~A.~Martins, E.~Menegoni, S.~Galli, G.~Mangano and A.~Melchiorri,
  Phys.\ Rev.\  D {\bf 82}, 023532 (2010)
  [arXiv:1001.3418 [astro-ph.CO]].

  
  \bibitem{ek}
  S.~Ekstrom, A.~Coc, P.~Descouvemont, G.~Meynet, \etal,
 Astron.\ Astrophys.\  {\bf 514}, 62 (2010),
  \url{arXiv:0911.2420 [astro-ph.SR]};\\
  A. Coc, \etal,
 Mem. Soc. Astron. Ital. {\bf80}, 658 (2009).
  
\bibitem{flam}
   V.~V.~Flambaum,
  \url{arXiv:physics/0302015};\\
   V.~V.~Flambaum and A.~F.~Tedesco,
  Phys.\ Rev.\  C {\bf 73}, 055501 (2006),
  \url{arXiv:nucl-th/0601050}.

\bibitem{lattice result ref 0402098}
   V.V.~Flambaum, D.B.~Leinweber, A.W.~Thomas, and R.D.~Young,
  Phys.\ Rev.\  D {\bf 69}, 115006 (2004),
  \url{arXiv:hep-ph/0402098}.
   
  \bibitem{clock-bize05}
 S. Bize, P. Laurent, M. Abgrall, H. Marion, \etal,
 J. Phys. B: At. Mol. Opt. Phys. {\bf38}, S449 (2005),
  \url{http://arXiv.org/abs/physics/0502117}.

\bibitem{nuclear radius effect}
  T.~H.~Dinh, A.~Dunning, V.~A.~Dzuba, V.~V.~Flambaum,
  Phys.\ Rev.\  {\bf A79}, 054102 (2009),
  [arXiv:0903.2090 [physics.atom-ph]].
  
\bibitem{BFK2011}
  J.~C.~Berengut, V.~V.~Flambaum, E.~M.~Kava,
  [arXiv:1109.1893 [physics.atom-ph]].

\bibitem{clock-prestage}
 J.D. Pretage, R.L. Tjoelker, and L. Maleki,
 Phys. Rev. Lett. {\bf 74}, 3511 (1995).

\bibitem{kappa-dzuba2}
 V.A. Dzuba, and V.V. Flambaum,
 Phys. Rev. A {\bf61}, 034502 (2001). 
 
\bibitem{kappa-dzuba3} 
 V.A. Dzuba, V.V. Flambaum, and M.V. Marchenko,
 Phys. Rev. A {\bf68}, 022506 (2003),
  \url{http://arXiv.org/abs/physics/0305066}. 
 
\bibitem{kappa-dzuba}
 V.A. Dzuba, V.V. Flambaum, and J.K. Webb,
 Phys. Rev. A {\bf59}, 230 (1999),
  \url{http://arXiv.org/abs/physics/9808021}. 
  
\bibitem{kappa-flamb}
 V.V. Flambaum,
 in {\em Laser Spectroscopy}, P. Hannaford, \etal Eds. (World Scientific, 2004) p. 47,
 \url{http://arXiv.org/abs/physics/0309107}.
 
  
\bibitem{clock-fisher04}
 M. Fischer, \etal,
 Phys. Rev. Lett. {\bf92}, 230802 (2004),
 \url{http://arXiv.org/abs/physics/0312086}.
  
\bibitem{clock-bize03}
 S. Bize,  S.A. Diddams, U. Tanaka, C.E. Tanner, \etal,
 Phys. Rev. Lett. {\bf90}, 150802 (2003),
  \url{ http://arxiv.org/abs/physics/0212109}.
  
\bibitem{clock-fortier07}
 T.M. Fortier, N. Ashby, J.C. Bergquist, M.J. Delaney, \etal,
 Phys. Rev. Lett. {\bf98}, 070801 (2007).

\bibitem{clock-peik04}
 E. Peik, B. Lipphardt, H. Schnatz, C. Tamm, \etal,
 Proc. of the 11th Marcel Grossmann meeting, Berlin, 2006,
 \url{http://arXiv.org/abs/physics/0611088}.

\bibitem{clock-peik06}
 E. Peik, B. Lipphardt, H. Schnatz, T. Schneider, \etal,
 Phys. Rev. Lett. {\bf93}, 170801 (2004),
 \url{http://arXiv.org/abs/physics/0402132}.
   
\bibitem{clock-blatt}
 S. Blatt,  A.D. Ludlow, G.K. Campbell, J.W. Thomsen, \etal,
 Phys. Rev. Lett. {\bf100}, 140801 (2008),
  \url{http://arXiv.org/abs/0801.1874}.

\bibitem{Dy-df}
 V.A. Dzuba, and V.V. Flambaum,
 Phys. Rev. A {\bf81}, 052515 (2010),
 \url{http://arxiv.org/abs/1003.1184}.
  
\bibitem{nguyen04}
 A.T. Nguyen, D. Budker, S.K. Lamoreaux, and J.R. Torgerson,
 Phys. Rev. A {\bf69}, 022105 (2004),
  \url{http://arxiv.org/abs/physics/0308104}.
  
\bibitem{clock-cingoz}
 A. Cing\"oz, A. Lapierre, A.-T. Nguyen, N. Leefer, \etal,
 Phys. Rev. Lett. {\bf98}, 040801 (2008),
 \url{http://arxiv.org/abs/physics/0609014}.
  
\bibitem{clock-rosen08}
 T. Rosenband, \etal,
 Science {\bf319}, 1808 (2008).       
 
 \bibitem{mu-theorie}
 R.I. Thompson,
 Astrophys. Lett. {\bf16}, 3 (1975).
 
\bibitem{clock-mu}
 A. Shelnikov, R.J. Butcher, C. Chardonnet, and A. Amy-Klein,
 Phys. Rev. Lett. {\bf100}, 150801 (2008),
 \url{http://arXiv.org/abs/0803.1829}.
    
\bibitem{constituent_bmm}
 See, for example, D.~Griffiths,
 {\it Introduction to elementary particles}
 (Weinheim, Germany: Wiley-VCH, 2008).

\bibitem{Fritzsch:2009xe}
  H.~Fritzsch and G.~Eldahoumi,
  Mod.\ Phys.\ Lett.\  A {\bf 24}, 2167 (2009),
  \url{arXiv:0906.1139 [hep-ph]}.  
  
  \bibitem{HY Cheng}
  H.~Y.~Cheng,
  Phys.\ Lett.\  B {\bf 219}, 347 (1989).

\bibitem{sigma term ref1}
   J.~R.~Ellis, A.~Ferstl, and K.~A.~Olive,
  Phys.\ Lett.\  B {\bf 481}, 304 (2000)
  \url{arXiv:hep-ph/0001005}.

\bibitem{sigma term ref2}
  J.~R.~Ellis, K.~A.~Olive, and C.~Savage,
  Phys.\ Rev.\  D {\bf 77}, 065026 (2008),
  \url{arXiv:0801.3656 [hep-ph]}.

  \bibitem{HF theorem ref}
  H.~Hellmann, Z.\ Phys. {\bf 85}, 180 (1933);\\
  R.~P.~Feynman,
  Phys.\ Rev.\  {\bf 56}, 340 (1939).

\bibitem{FH Gasser ref}
   J.~Gasser,
  Annals Phys.\  {\bf 136}, 62 (1981).

\bibitem{trace anomaly ref}
  R.~J.~Crewther,
  Phys.\ Rev.\ Lett.\  {\bf 28}, 1421 (1972);\\
  M.~S.~Chanowitz and J.~R.~Ellis,
  Phys.\ Lett.\  B {\bf 40}, 397 (1972);\\
  M.~S.~Chanowitz and J.~R.~Ellis,
  Phys.\ Rev.\  D {\bf 7}, 2490 (1973);\\
  S.~L.~Adler, J.~C.~Collins, and A.~Duncan,
  Phys.\ Rev.\  D {\bf 15}, 1712 (1977);\\
  J.~C.~Collins, A.~Duncan, and S.~D.~Joglekar,
  Phys.\ Rev.\  D {\bf 16}, 438 (1977);\\
  N.~K.~Nielsen,
  Nucl.\ Phys.\  B {\bf 120} (1977) 212.
   
\bibitem{SVZ}
  M.~A.~Shifman, A.~I.~Vainshtein, and V.~I.~Zakharov,
  Phys.\ Lett.\  B {\bf 78}, 443 (1978).  

  
\bibitem{Li and Cheng}  
  L.~F.~Li and T.~P.~Cheng,
  \url{arXiv:hep-ph/9709293}.  
  
 \bibitem{Gasser and Leutwyler quark masses}
   J.~Gasser and H.~Leutwyler,
  Phys.\ Rept.\  {\bf 87}, 77 (1982).

 
  \bibitem{Leutwyler 96}
   H.~Leutwyler,
  Phys.\ Lett.\  B {\bf 378}, 313 (1996),
  \url{arXiv:hep-ph/9602366}.

  
  
\bibitem{sigma_0 ref}
  B.~Borasoy and U.~G.~Meissner,
  Annals Phys.\  {\bf 254}, 192 (1997),
  \url{arXiv:hep-ph/9607432}.  
  
  \bibitem{morespn}
   J.~Gasser, H.~Leutwyler, and M.~E.~Sainio,
  Phys.\ Lett.\  B {\bf 253}, 252 (1991);\\
   M.~Knecht,
  PiN Newslett.\  {\bf 15}, 108 (1999),
  \url{arXiv:hep-ph/9912443};\\
  M.~E.~Sainio,
  PiN Newslett.\  {\bf 16}, 138 (2002),
  \url{arXiv:hep-ph/0110413}.
  
  \bibitem{PDG ref}
   K.~Nakamura, \etal,  [Particle Data Group],
  J.\ Phys.\ G {\bf 37}, 075021 (2010).

\bibitem{HK ref}
  T.~Kunihiro and T.~Hatsuda,
  Phys.\ Lett.\  B {\bf 240}, 209 (1990);\\
  T.~Hatsuda and T.~Kunihiro,
  Z.\ Phys.\  C {\bf 51} (1991) 49;\\
  T.~Hatsuda and T.~Kunihiro,
  Nucl.\ Phys.\  B {\bf 387}, 715 (1992). 



\bibitem{De Rujula:1975ge}
 A.~De Rujula, H.~Georgi, and S.~L.~Glashow,
  Phys.\ Rev.\  D {\bf 12}, 147 (1975).
     
\bibitem{Manohar and Georgi}
  A.~Manohar and H.~Georgi,
  Nucl.\ Phys.\  B {\bf 234}, 189 (1984).
  
\bibitem{chiQM ref}
  L.~Y.~Glozman and D.~O.~Riska,
  Phys.\ Rept.\  {\bf 268}, 263 (1996),
  \url{arXiv:hep-ph/9505422}.
     
\bibitem{Isgur:1978xj}
  N.~Isgur and G.~Karl,
  Phys.\ Rev.\  D {\bf 18}, 4187 (1978);\\
  N.~Isgur and G.~Karl,
  Phys.\ Rev.\  D {\bf 19}, 2653 (1979)
  [Erratum-ibid.\  D {\bf 23}, 817 (1981)].
  

  
\bibitem{chiSSB ref}
   J.~T.~Goldman and R.~W.~Haymaker,
  Phys.\ Rev.\  D {\bf 24}, 724 (1981);\\
  M.~K.~Volkov,
  Annals Phys.\  {\bf 157} (1984) 282;\\
  D.~Ebert and H.~Reinhardt,
  Nucl.\ Phys.\  B {\bf 271} (1986) 188.
     
\bibitem{NJL ref}
   Y.~Nambu and G.~Jona-Lasinio,
  Phys.\ Rev.\  {\bf 122}, 345 (1961);\\
  Y.~Nambu and G.~Jona-Lasinio,
  Phys.\ Rev.\  {\bf 124}, 246 (1961).

\bibitem{NJL three flavor ref}
   S.~Klimt, M.~F.~M.~Lutz, U.~Vogl, and W.~Weise,
  Nucl.\ Phys.\  A {\bf 516}, 429 (1990);\\
  U.~Vogl, M.~F.~M.~Lutz, S.~Klimt, and W.~Weise,
  Nucl.\ Phys.\  A {\bf 516}, 469 (1990).
     
\bibitem{relativisitic corrections ref}
   H.~Georgi and A.~Manohar,
  Phys.\ Lett.\  B {\bf 132}, 183 (1983).

\bibitem{chipt classical}
  S.~Weinberg,
  Physica A {\bf 96}, 327 (1979);\\
  J.~Gasser and H.~Leutwyler,
  Annals Phys.\  {\bf 158}, 142 (1984);\\
  J.~Gasser and H.~Leutwyler,
  Nucl.\ Phys.\  B {\bf 250}, 465 (1985);\\
  J.~Gasser, M.~E.~Sainio, and A.~Svarc,
  Nucl.\ Phys.\  B {\bf 307}, 779 (1988);\\
  A.~Krause,
  Helv.\ Phys.\ Acta {\bf 63} (1990) 3.
  
\bibitem{chipt pedagogical review}
   For a pedagogical review, see, for example, S.~Scherer,
  Adv.\ Nucl.\ Phys.\  {\bf 27}, 277 (2003),
  \url{arXiv:hep-ph/0210398}.
   
\bibitem{chipt LECs}
   V.~Bernard,
  Prog.\ Part.\ Nucl.\ Phys.\  {\bf 60}, 82 (2008),
  \url{arXiv:0706.0312 [hep-ph]}.
  
\bibitem{dim analysis}
    H.~Georgi and L.~Randall,
  Nucl.\ Phys.\  B {\bf 276}, 241 (1986);\\
  H.~Georgi,
  Phys.\ Lett.\  B {\bf 298}, 187 (1993),
  \url{arXiv:hep-ph/9207278}.
  
\bibitem{HBchiPT ref}
  E.~E.~Jenkins and A.~V.~Manohar,
  Phys.\ Lett.\  B {\bf 255}, 558 (1991).
  
\bibitem{HBchiPT mm ref1}
  E.~E.~Jenkins, M.~E.~Luke, A.~V.~Manohar, and M.~J.~Savage,
  Phys.\ Lett.\  B {\bf 302}, 482 (1993)
  [Erratum-ibid.\  B {\bf 388}, 866 (1996)],
  \url{arXiv:hep-ph/9212226}.

\bibitem{HBchiPT mm ref2}
   U.~G.~Meissner and S.~Steininger,
  Nucl.\ Phys.\  B {\bf 499}, 349 (1997),
  \url{arXiv:hep-ph/9701260}.
  
\bibitem{HBchiPT mm ref3}
    L.~Durand and P.~Ha,
  Phys.\ Rev.\  D {\bf 58}, 013010 (1998),
  \url{arXiv:hep-ph/9712492}.
 
\bibitem{HBchiPT mm ref4}
   S.~J.~Puglia and M.~J.~Ramsey-Musolf,
  Phys.\ Rev.\  D {\bf 62}, 034010 (2000),
  \url{arXiv:hep-ph/9911542}.
  
\bibitem{EOMS ref}
   T.~Fuchs, J.~Gegelia, G.~Japaridze, and S.~Scherer,
  Phys.\ Rev.\  D {\bf 68}, 056005 (2003),
  \url{arXiv:hep-ph/0302117}.
  
\bibitem{EOMS mm ref1}
   L.~S.~Geng, J.~Martin Camalich, L.~Alvarez-Ruso, and M.~J.~Vicente Vacas,
  Phys.\ Rev.\ Lett.\  {\bf 101}, 222002 (2008),
  \url{arXiv:0805.1419 [hep-ph]}.

\bibitem{EOMS mm ref2}
   L.~S.~Geng, J.~Martin Camalich, and M.~J.~Vicente Vacas,
  Phys.\ Lett.\  B {\bf 676}, 63 (2009),
  \url{arXiv:0903.0779 [hep-ph]}.
    
\bibitem{Pade lattice}
   D.~B.~Leinweber, D.~H.~Lu, and A.~W.~Thomas,
  Phys.\ Rev.\  D {\bf 60}, 034014 (1999),
  \url{arXiv:hep-lat/9810005};\\
  E.~J.~Hackett-Jones, D.~B.~Leinweber and A.~W.~Thomas,
  Phys.\ Lett.\  B {\bf 489}, 143 (2000),
  \url{arXiv:hep-lat/0004006}.
  
\bibitem{FFR ref}
   J.~F.~Donoghue, B.~R.~Holstein, and B.~Borasoy,
  Phys.\ Rev.\  D {\bf 59}, 036002 (1999)
  \url{arXiv:hep-ph/9804281};\\
  D.~B.~Leinweber, A.~W.~Thomas, and R.~D.~Young,
  Phys.\ Rev.\ Lett.\  {\bf 92}, 242002 (2004),
  \url{arXiv:hep-lat/0302020}.
   
\bibitem{x21cm}
  R. Srianand, N. Gupta, P. Petitjean, P. Noterdaeme, \etal,
  Mon. Not. R. Astron. Soc. {\bf405}, 1888 (2010),
  \url{arXiv:1002.4620}.
  
\bibitem{y-murph}
  M.T. Murphy, J.K. Webb, V.V. Flambaum, M.J. Drinkwater, \etal,
  Mon. Not. R. Astron. Soc. {\bf 327}, 1244 (2001),
  \url{astro-ph/0101519}.

\bibitem{OH-3}
 N. Kanekar, C.L. Carilli, G.I. Langston, G. Rocha, \etal,
 Phys. Rev. Lett. {\bf 95}, 261301 (2005),
 \url{astro-ph/0510760}.

\bibitem{darling}
  J. Darling, 
  Astrophys. J. {\bf612}, 58 (2004),
  \url{astro-ph/0405240}.

\bibitem{OH-1}
  J.N. Chengalur, and N. Kanekar,
  Phys. Rev. Lett. {\bf91}, 241302 (2003),
  \url{astro-ph/0310764}.
   
\bibitem{kanekar1}
 N. Kanekar, J.N. Chengalur, and T. Ghosh,
 Phys. Rev. Lett. {\bf93}, 051302 (2004),
 \url{arXiv:astro-ph/0406121}.

\bibitem{kanekar1bis}
 N. Kanekar, {\em et al.},
 Astrophys. J. {\bf716}, L23 (2010).
  
\bibitem{kanekar2}
   N. Kanekar,
    Mod. Phys. Lett. A {\bf23}, 2711 (2008),
    \url{arXiv:0810.1356}.
  
\end{thebibliography}
\end{document}